\documentclass[unstructuredabstract]{aa}  

\usepackage{graphicx}
\usepackage{txfonts}
\usepackage{lipsum}
\usepackage{subcaption}         
\usepackage{lscape}             
\usepackage{placeins}           
\usepackage{cleveref}  
\usepackage{xcolor}
\usepackage[normalem]{ulem}
\newcommand{\kms}{\ensuremath{\,\mathrm{km\,s^{-1}}}}
\begin{document}

   \title{Proximate damped Lyman-$\alpha$ systems as tracers of quasar feedback  
}


   \author{Patrick Petitjean\\
        }
   \institute{Institut d'Astrophysique de Paris and Sorbonne Université, 98bis Boulevard Arago,
                75014, France\\
             \email{petitjean@iap.fr}\\
}

   \date{Received XX, 20XX}

\abstract{Active galactic nuclei (AGN) profoundly affect the interstellar medium of their host galaxies through intense radiation fields and powerful winds. Characterising this feedback is essential for understanding galaxy formation and evolution.
Here we revisit the origin of proximate damped Lyman-$\alpha$ absorbers (PDLAs), which trace cold gas within 3000~\kms\ of the quasar redshift, and interpret their kinematics and physical properties within a unified framework.
We searched the SDSS DR16 database for low-ionisation metal absorption-line systems 
at the quasar redshift (referred to as ProxSys).
This approach enables us to identify and classify different types of proximate absorbers, including so-called Ghostly systems, coronagraphic DLAs (DLA-Cor), standard DLAs, and sub-DLAs, based on the presence of strong Lyman-$\alpha$ absorption, partial covering signatures, or excited atomic transitions such as Si~{\sc ii}$^*$.
We find that about 13\% of  ProxSys belong to the Ghostly or DLA-Cor classes and exhibit strong absorption from excited species. The different classes of ProxSys 
form a continuous sequence characterised by decreasing Si~{\sc ii}$^*$, C~{\sc iv}, and N~{\sc v} absorption strengths and dust content. Their velocity distributions reveal multiple kinematic components. Standard DLAs cluster within $\pm1000$~\kms\ of the quasar systemic redshift, consistent with gas in the host galaxy, whereas Ghostly and Si~{\sc ii}$^*$-bearing systems display broader distributions, including outflows reaching $-2000$~\kms\ and a smaller population of inflowing clouds up to $+1200$~\kms. Median stacked spectra confirm that Ghostly and coronagraphic systems arise in dense, compact gas partially covering the quasar emission regions.
These results support a scenario in which cold, dense clouds participate in a dynamic cycle of inflow and outflow in the vicinity of quasars, consistent with chaotic cold accretion.

 } 
 

   \maketitle
\nolinenumbers

\section{Introduction}
 
Bright quasars at high-redshift are luminous sources powered by the accretion of material from their host galaxies onto a massive black hole. Their intense emission, together with the 
launching of powerful winds, affects the interstellar medium of the galaxy, either inhibiting or promoting star formation \citep[e.g.][]{Dubois2013}. This feedback is thought to play a major role in galaxy evolution by quenching or enhancing star-formation
\citep[e.g.][]{Springel2005,Cattaneo2009,Schaye2015,Dubois2016,Piotrowska2022,Bluck2023,Lauzikas2024}.

Active galactic nuclei (AGN) show evidence for multi-phase, multi-scale winds covering a wide range of ionisation states and spatial scales 
\citep{Laha2021}.
In X-rays, a variety of absorption features trace highly ionised, fast outflows. 
Soft X-ray ``warm absorbers'' are common, detected in roughly $\sim$50\% of radio-quiet AGN \citep{Laha2014}, 
and typically exhibit velocities of a few hundred up to a few thousand km~s$^{-1}$. 
More dramatically, ultra-fast outflows (UFOs) are revealed through blue-shifted Fe~K-shell absorption lines, indicating highly ionised gas accelerated to
relativistic velocities of $\sim$0.1--0.3~c. 
As many as $\sim$40\% of nearby AGN show evidence for UFOs originating on sub-parsec scales \citep{Tombesi2011, Gofford2015}, suggesting that powerful, highly ionised winds are a relatively common feature of accreting supermassive black holes.

In the UV and optical, 15 to 20\% of AGNs exhibit blue-shifted high ionisation absorption lines (C~{\sc iv}, N~{\sc v}, O~{\sc vi}...)  often in the form of Broad Absorption Line (BAL) systems \citep{Weymann1981,Gibson2009}, with outflow velocities up to a few 10$^{3-4}$~\kms. The variations of the absorptions 
 \citep{FilizAk2012, Aromal2024} argue for the gas being located close to
 the accretion disc although ionisation arguments yield to much larger distances
 \citep{Arav2020}.

BAL winds are radiatively driven, and the enhanced UV reddening of BAL quasars
relative to non-BAL quasars indicates that both line absorption and dust
radiation pressure contribute to accelerating the outflow \citep{Laor2002}.  
\citet{Murray1995} explain that the inner edge of the disc wind 
giving rise to BALs has the same characteristics as a ``warm absorber". 
\citet{Tombesi2013} conclude that X-ray UFOs and the comparatively lower
velocity warm absorbers could represent parts of a single
large-scale stratified outflow observed at different locations from the
black hole. Actually, although a distinction between the outflows is made because
they are detected with different techniques, they may share similar origin although not located at the same distance to the central AGN 
\citep[e.g.][]{Bu2021}.

AGN ionised winds are observed on much larger scales \citep{Zakamska2016, Leung2019, ForsterSchreiber2019}.  Outflows are detected in 17\% of the AGNs in the redshift range $1.4\le z \le 3.8$ with spatial extents of 0.3-11~kpc and velocities in the range $400-3000$~\kms. \citet{Temple2024} have  found a weak correlation between 
the velocities of the [O III] emission and associated C~{\sc iv} BALs. 
They conclude that this is consistent with a scenario wherein BAL troughs are intermittent tracers of persistent quasar outflows.

Neutral and molecular outflows are commonly seen from nearby AGNs \citep{Rupke2021} to high-redshifts \citep{Davies2024}. 
They are thought to be the consequence of the action of AGN winds onto
the ISM of the host-galaxy. They can be traced from the inner region of
the AGN in the radio \citep{Morganti2020} to large scales \citep{Belli2024}.
However how these large scale outflows that may quench star
formation in the galaxy are launched and how radiation and 
particle outflows couple with the host galaxy gas is still a matter of debate. 

Disc wind outflows are the most efficient way to deposit energy 
in and out of the AGN host-galaxy. Current theoretical understanding of active galactic nucleus outflows tells that 
a constant velocity momentum-driven wind is rapidly accelerated 
following inefficient Compton cooling of post-shock material \citep{Marconcini2025}. 
Indeed, it seems probable that outflows are energy conserving. 
Although this has been demonstrated for fast launching winds ($v \sim 10,000$\kms), slower winds ($v \sim 1000$\kms) are also possible \citep{Faucher2012}. 

Winds and radiation from the central AGN compress and stir the neutral gas in and around the host galaxy. It is therefore expected that some signature of this interaction should be imprinted in the quasar spectrum as H~{\sc i} absorption, in particular as proximate damped Lyman-alpha systems (DLAs) with small velocities relative to the quasar (typically $\Delta V < 3000$\kms).

\citet{Ellison2002} searched for DLAs in a radio-selected QSO sample and found that the density of proximate DLAs (PDLAs) is about four times higher than that of intervening systems. No excess of proximate DLAs was found in a radio-quiet sample, however. This result was partially confirmed by \citet{Prochaska2008}, who reported that the density of proximate DLAs is roughly twice that of intervening DLAs in the redshift range $2.5 < z < 3.5$, and otherwise consistent. However considering that
the observed enhancement of PDLAs around quasars is the result of quasar-galaxy clustering they estimate that the number of proximate DLAs is smaller by a factor 
of 5 to 10 than expected, interpreting this as the result of the quasar ionising flux photo-evaporating H~{\sc i} in the host and nearby galaxies. 

Later, \citet{Ellison2010,Ellison2011} studied, respectively, seven proximate DLAs observed at high resolution and 85 systems from SDSS DR5. They found that proximate DLAs generally have higher metallicities than intervening absorbers, and interestingly, that absorbers near QSOs with lower rest-frame UV luminosities exhibit significantly stronger metal lines. They speculate that absorbers near high-luminosity QSOs may have experienced premature quenching of star formation. H~{\sc i} has also been observed in absorption in the halos of bright quasars within 100~kpc \citep{Prochaska2009}.

It is however known that the clouds giving rise to proximate DLAs can be 
sufficiently small that they do not fully cover the narrow-line region 
\citep[so-called coronagraph DLAs or DLA-Cor,][]{Hennawi2009,Finley2013,Jiang2016,Pan2018,Xie2018,Balashev2020}, 
or even the broad-line region 
\citep[so-called Ghostly DLA,][]{Fathivavsari2017}. 
These systems have been observed across a wide range of quasar types \citep{Wampler1995,Dai2020,Gupta2022,Wu2024,Rojas2025}.
The corresponding clouds have been 
claimed to be located close to the AGN \citep{Laloux2021}. Recently
\citet{Noterdaeme2019} observed an excess of H$_2$ bearing PDLAs. 
They conclude that although most of the systems
probably originate from galaxies in the quasar group, a small fraction 
of them could be located in the quasar host itself.

Here, we revisit the nature of proximate DLAs in light of recent observations and 
simulations, substantially increasing the available statistics thanks to the SDSS DR16 dataset. In Section~2, we describe our procedure to detect proximate systems and classify
them into different categories. Section~3 presents the number of systems, while 
Section~4 shows the median spectra of the various classes. In Section~5, we examine 
their velocity distributions, and Section~6 discusses
the properties of the quasars. We interpret our findings in Section~7 and summarise our
conclusions in Section~8.

\section{The procedure}
\subsection{Definitions}

We aim to investigate the presence of gas with high 
H~{\sc i} column density in the vicinity of the active nucleus. 
Such gas is usually detected through its H~{\sc i} Lyman-$\alpha$ 
absorption signature 
in the quasar spectrum at relative velocities less than about 
3000\kms\ with respect to the quasar redshift. 
We refer to these absorption systems as proximate systems (ProxSys)
in general. Among these, we call, as usual in the literature, 
proximate DLAs (PDLA)  systems with 
H~{\sc i} column density log~$N$(H~{\sc i})~$>$~20.3. 
These systems have been searched for in several SDSS releases, 
and corresponding catalogues are available \citep{noterdaeme2012, chabanier2022}.

However, since our work relies on SDSS data, it 
is difficult to impose 
such an arbitrary column density threshold because the low resolution and 
signal-to-noise ratio of SDSS spectra make damping wings of the 
Lyman-$\alpha$ trough hard to detect. Also
in such spectra only lower limits on the equivalent width can be derived. 
We therefore do not restrict our sample to DLAs alone, and also include 
sub-DLAs. 
In addition, ProxSys include Ghostly and DLA-Cor systems that cannot
be detected by their Lyman-\(\alpha\) trough.

If we wish to probe the full population of ProxSys down to H~{\sc i} column densities close to, or even somewhat 
smaller than those of PDLAs, a different selection strategy is required, which
does not use the presence of a strong H~{\sc i} Lyman-$\alpha$ trough.

We thus select all absorption systems that can be identified through strong metal lines, without reference to their H~{\sc i} content. Although this is beyond the scope of the present paper, this approach may allow us to distinguish between gas expelled by the AGN, which is expected to be metal-rich, and gas originating from the IGM, presumably of lower metallicity.

\subsection{Search for metallic ProxSys}

We use the final SDSS-IV quasar catalogue from Data Release~16 (DR16) of the extended Baryon Oscillation Spectroscopic Survey \citep[eBOSS;][]{Dawson2016, Ahumada2020}, which we will refer to as DR16Q. This catalogue contains 750,414 quasars \citep{Lyke2020}.

We select quasars with quasar redshift, absorption index (characterising the
strength of BALs) and median signal-to-noise ratio,
$z_{\rm em}$~$>$~2.11, AI\texttt{\_}CIV~$<$~1000\kms, 
and SN\texttt{\_}MEDIAN\texttt{\_}ALL~$>$~1.95.
We use the best available redshift as reported on line~27 of Table~D1 in \citet{Lyke2020}. The minimum redshift threshold is chosen so that we can estimate the strength of the Lyman-$\alpha$ absorption associated with the ProxSys. 
The C~{\sc iv} absorption index (AI\texttt{\_}CIV) is used to avoid broad absorption lines. The minimum SNR is set arbitrarily, based on a few visual inspections, to avoid very noisy spectra and minimise false positives. After applying these criteria, 
we are left with a total of 164,901 quasars.

We first applied a simple algorithm to select metal line systems, summing 
the velocity windows from $-$3000 to $+$3000~\kms~ around the
rest wavelengths of the transitions Si~{\sc ii}1260, Si~{\sc ii}$^*$1264, 
C~{\sc ii}1334, Al~{\sc ii}1670 at the redshift of the quasar
and derive the position of a minimum in the resultant spectrum. We then calculated 
the equivalent widths and corresponding errors for 
all transitions above 1200~\AA~ at the redshift of this position
and selected systems with at least
two high ionisation transitions (from the N~{\sc v} and C~{\sc iv} doublets)
and two low-ionisation transitions (from the Si~{\sc ii}1260, Si~{\sc ii}$^*$1264,
O~{\sc i}1302, C~{\sc ii}1334 and Al~{\sc ii}1670 transitions) detected at the 3~$\sigma$ level.
These transitions often lie in the wavelength regions of highest SNR.
The Mg~{\sc ii} doublet is not used at this stage because a number of 
spectra are spoiled in the corresponding wavelength region by inadequate 
sky subtraction. We use it during the confirmation process.
The required number of detected transitions was chosen to  minimise
false positives. Note however, that a few H$_2$ bearing 
proximate systems without detected high ionisation transitions are known
\citep{Noterdaeme2019}. These systems are indeed interesting and could be studied separately, but they are rare.
Following this procedure we detected 5373 metal ProxSys candidates.

\subsection{Visual inspection}

To confirm the reality of the ProxSys, we visually inspected the
5373 spectra selected as described in the previous section.
A total of 1071 systems turned out either to be fakes (940) or could not be firmly classified (131), and 45 quasars were identified as BALs missed by DR16.
For the remaining 4257 confirmed metal-line ProxSys, we measured the equivalent width of the associated H~{\sc i} absorption when present. Given the spectral resolution and SNR of the spectra, this measurement is highly uncertain, especially since the absorption lies on top of the quasar Lyman-$\alpha$ emission. We therefore arbitrarily 
classified the systems as HighWSys, LowWSys and VeryLowWSyst
 as those with $W$(Lyman$\alpha$)~$>$~5, between 2 and 5, and $<$~2~\AA, respectively. 

We then sorted the systems into several classes:
(i) HighWSys or HighWSys?, (ii) LowWSys or LowWSys?, (iii) VeryLowWSys or VeryLowWSys?, where the question mark indicates doubtful cases, mostly due to blending.
We also defined more specific classes:
(iv) DLA-Cor and DLA-Cor?, corresponding to coronagraphic DLAs when a narrow emission 
line is present at the bottom of a detected DLA trough; the question mark indicates systems where no emission line is seen but the trough does not reach zero flux, indicating some leakage;
(v) Ghost and Ghost?, referring to systems where no or very little absorption is detected at the position of the Lyman-$\alpha$ line, with the question mark indicating ambiguous cases due to low SNR or slightly shifted absorption relative to the metal lines. 
In addition to these classes, we defined an additional class as ProxSys*
gathering all systems from the above classes where absorption from  
Si~{\sc ii}$^*$$\lambda$1264 is clearly detected. Other absorption lines
are usually too weak to be detected in individual spectra and C~{\sc ii}$^*$$\lambda$1335 is
always badly blended with C~{\sc ii}$\lambda$1334.
The selection has been performed with great care, but some low-level contamination by 
interlopers may still be present.
 
Particular attention was given to Ghost systems and a second visual inspection was performed.
Among the 220 systems classified as Ghost (90) or Ghost? (130), we confirmed 95 Ghost systems and 94 probable Ghosts, and rejected 31. Since these systems are of particular interest for characterising the spatial distribution of the Broad Line Region 
\citep[see][]{Laloux2021}, we will report their properties in more details in a forthcoming paper.
For the other classes, we find 1688 (resp. 60) HighWSys (resp. HighWSys?),
227 (resp. 96) DLA-Cor (resp. DLA-Cor?),
1554 (resp. 250) LowWSys (resp. LowWSys?), and
119 (resp. 43) VeryLowWSys (resp. VeryLowWSys?).

Note that Ghost and DLA-Cor systems have been shown to be associated with DLAs. 
Out of a total of 3873 ProxSys, we have identified 189 highly probable Ghosts and 323 DLA-Cor systems. Together, these two categories therefore account for about 13\% 
of the metal-rich ProxSys.

\begin{figure}[h]
   \centering
   \includegraphics[width=\linewidth, keepaspectratio]{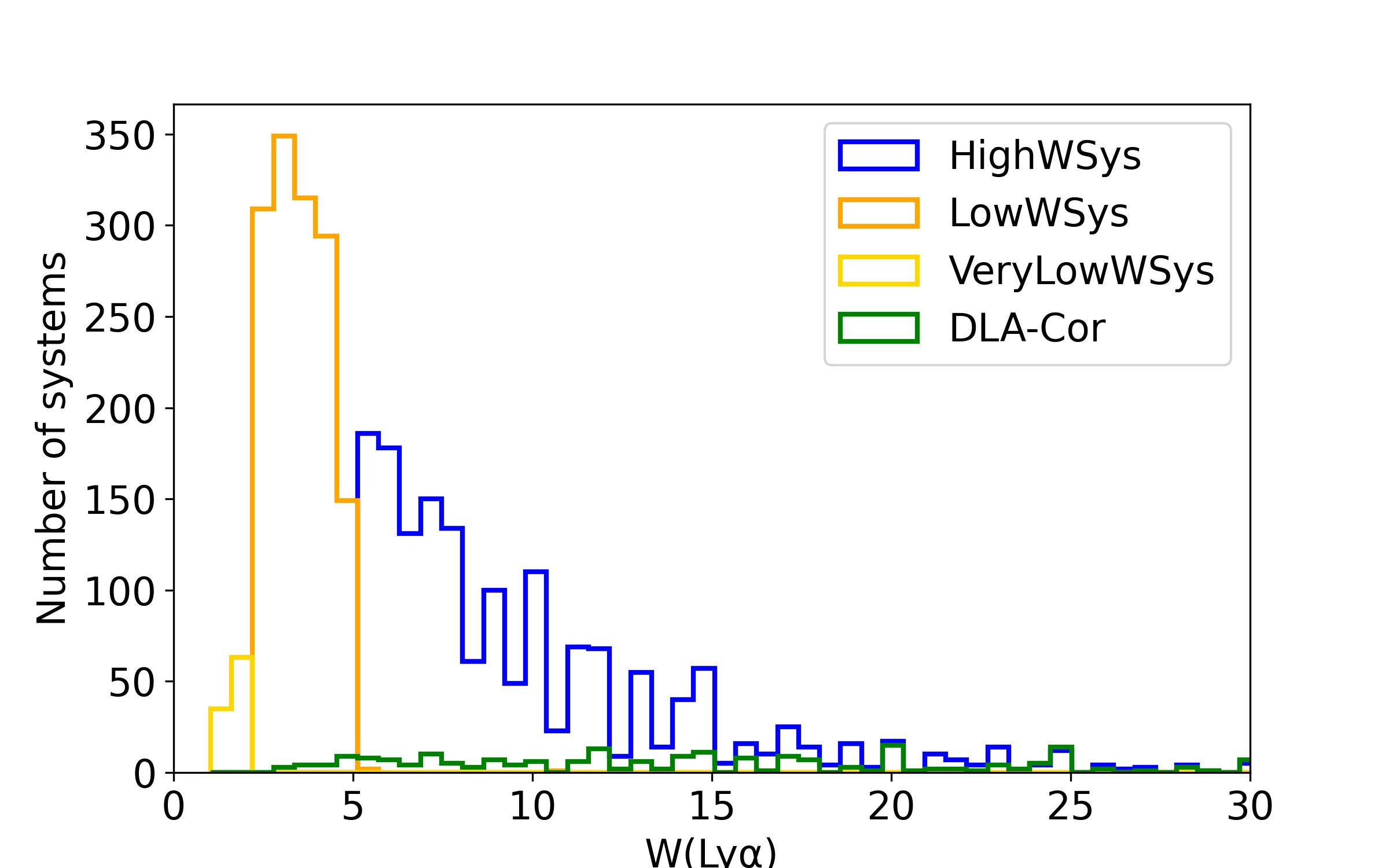}
      \caption{Distributions of the H~{\sc i} Lyman$\alpha$ absorption line 
      equivalent width for HighWSys (blue), LowWSys (orange), VeryLowWSys (yellow) and DLA-Cor (green) as defined in the text.}
         \label{Wdist}
   \end{figure}

For each detected system, we automatically measured the equivalent width ($W$) of the Lyman-$\alpha$ absorption line. The distributions of $W$ for HighWSys, LowWSys, VeryLowWSys and DLA-Cor are shown in Fig.~\ref{Wdist}.
We caution that this measurement is only a rough estimate of $W$, because we measure solely the central portion of the absorption trough, which lies within the wing of the Lyman-$\alpha$ emission line. The resulting values should therefore be considered lower limits on the true effective equivalent width.
It is interesting to note that, although the probability of coronagraph DLAs having large equivalent widths is higher compared to the overall ProxSys distribution, they are nevertheless observed across the full range of equivalent widths.

We checked whether we recover the Ghostly DLAs detected by \citet{Fathivavsari2020}. Out of the 89 systems, we recover 58. Of the 31 missing systems, seven (7) are misidentified Ghostly DLAs, nine (9) are questionable (including three with 1000$>$AI$>$900~km/s), and nine (9) lie outside our search range (four with AI$>$1000~km/s and five with $z$$<$2.11). Therefore, we miss six confirmed Ghostly DLAs, yielding a success rate of 86\%. Among these six, three have no C~{\sc ii} line detected but are confirmed by strong Mg~{\sc ii} absorption, while three show no high-ionisation absorption lines. This explains why these systems were missed. On the other hand, we find 36 additional confirmed Ghostly DLAs and 94 probable ones. A detailed analysis of these systems will be presented in a future paper.

Out of the 81 systems in \citet{Noterdaeme2019} consisting of systems detected solely 
via the presence of molecular hydrogen, apart from those outside our selection, we miss
nine (9), four of which showing no metals at all.
PDLAs without metals or high-ionisation associated lines although rare are of particular
interest and may be studied in future work. 

We thus conclude that we recover most of the systems already known
but more importantly detect twice as much Ghostly DLAs.

\section{Number of metal ProxSys}
\subsection{Comparison with PDLAs}
We aim to compare Metal ProxSys systems with PDLAs.
To do so, we require a catalogue of PDLAs identified through their Lyman-$\alpha$ absorption trough. 
PDLAs have been searched for in SDSS-DR5 by \citet{Prochaska2008} (for log$N$(H~{\sc i})~$>$~20.3) and in SDSS-DR9 and SDSS-DR16 by, respectively, \citet{noterdaeme2012} and
\citet{chabanier2022} (for log$N$(H~{\sc i})~$>$~20). 

\citet{chabanier2022} provide a catalogue of PDLAs which appears at first sight to be 
suitable for our purpose.
However, we realised that their DLA search was performed only up to the quasar emission redshift, $z_{\rm em}$, whereas PDLAs can be detected with $z_{\rm abs} > z_{\rm em}$. Keeping this limitation in mind and because it is derived from the same data release as our work, we nevertheless make use of their catalogue for the present comparison.

\begin{figure}[h]
   \centering
   \includegraphics[width=\linewidth, keepaspectratio]{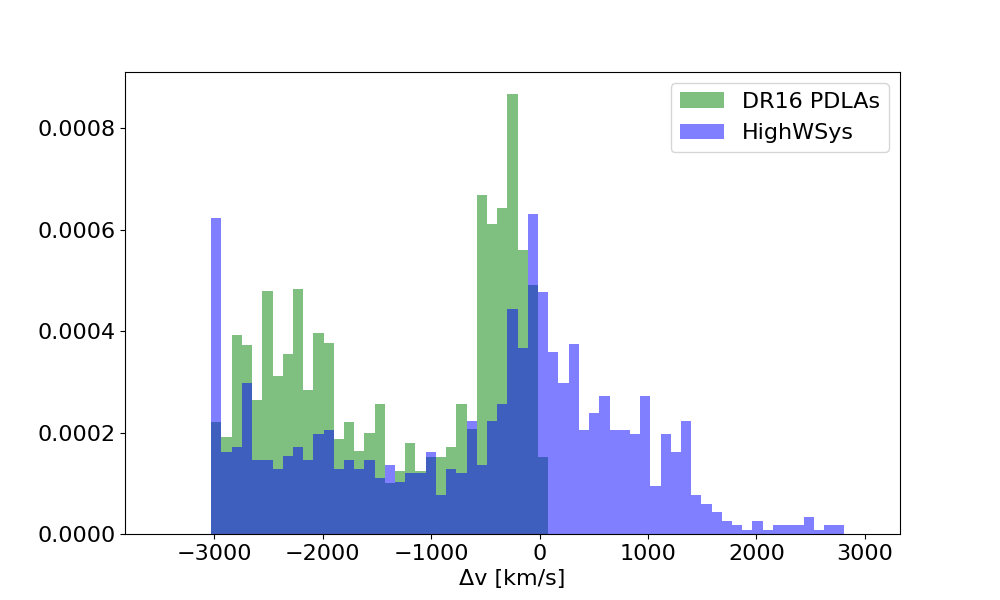}
      \caption{Distributions of the velocity difference between the 
      system and the quasar for the HighWSys ProxSys (blue) and the
      PDLAs in \cite{chabanier2022} (green).}
         \label{HistodeltavPDLA}
   \end{figure}

In Fig.~\ref{HistodeltavPDLA}, we plot the velocity of the systems relative to the quasar emission redshift for the PDLAs in DR16 with log~$N$(H~{\sc i})~$>$~20, as reported by \citet{chabanier2022}, together with our HighWSys sample. 
It is apparent that PDLAs with $z_{\rm abs}$~$>$~$z_{\rm em}$ are not included in 
the Chabanier's sample. 
It is worth noting on Fig.~\ref{HistodeltavPDLA} that the shapes of the distributions are
similar (see below however). In addition there is a peak at $z_{\rm em}$
and a minimum around $-$1000\kms which stayed unnoticed in previous studies.

\begin{figure}[h]
   \centering
\includegraphics[width=0.9\columnwidth]{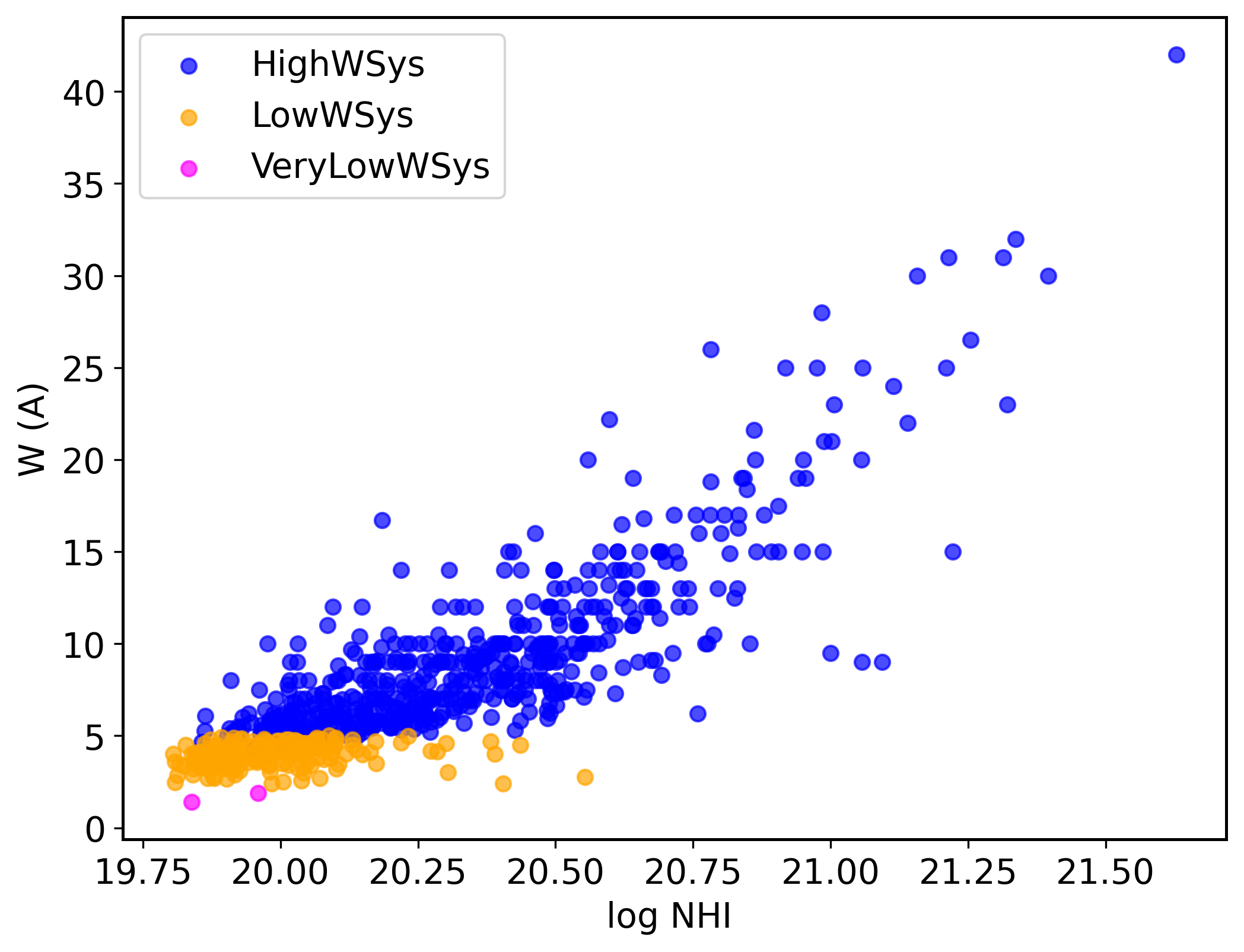}
      \caption{Equivalent 
width of the Lyman-$\alpha$ trough measured for our systems versus the logarithm of the H~{\sc i} column density as measured by
\citet{chabanier2022}. The blue, orange and red points correspond to
High, Low and Verylow-W systems. }
\label{WvsNHI}
   \end{figure}

\begin{figure}[h]
   \centering
\includegraphics[width=0.9\columnwidth]{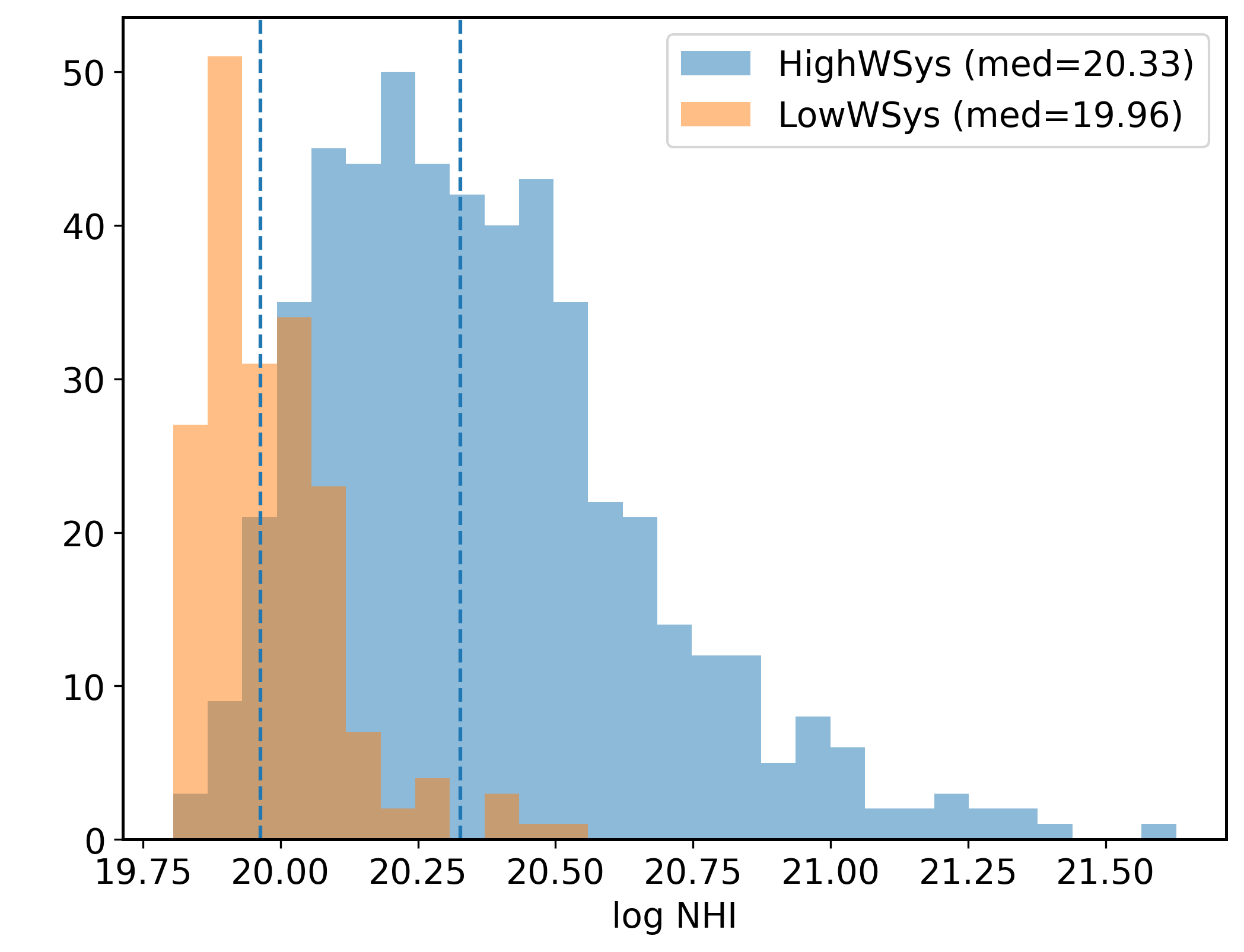}
      \caption{Distributions of the H~{\sc i} column densities for 
      High- (blue) and Low- (orange) Wsyst  for the 
      770 systems common to our and Chabanier's samples.}
         \label{HistoNHI}
   \end{figure}

In Fig.~\ref{WvsNHI} we plot the equivalent 
width of the Lyman-$\alpha$ trough measured for our systems versus the logarithm of the H~{\sc i} column density as measured by
\citet{chabanier2022}. The blue, orange and red points correspond to
High, Low and Verylow-W systems respectively. 
There are 710 systems common to our and Chabanier's samples, 524, 184 and 2 
High, Low and Verylow-W systems, respectively. 
It is clear that most of our systems are bona fide DLAs or sub-DLAs
and that the corresponding equivalent widths are underestimated.
The distributions of  log~$N$(H~{\sc i}) for these systems as measured by \citet{chabanier2022} are given in Fig.~\ref{HistoNHI}.


The fact that we recover 710 systems from the Chabanier's catalogue implies that
we recover a significant fraction ($>$20\%) of bona-fide PDLAs.

\subsection{Distribution with redshift}
\begin{figure}[h]
   \centering
   \includegraphics[width=\linewidth, keepaspectratio]{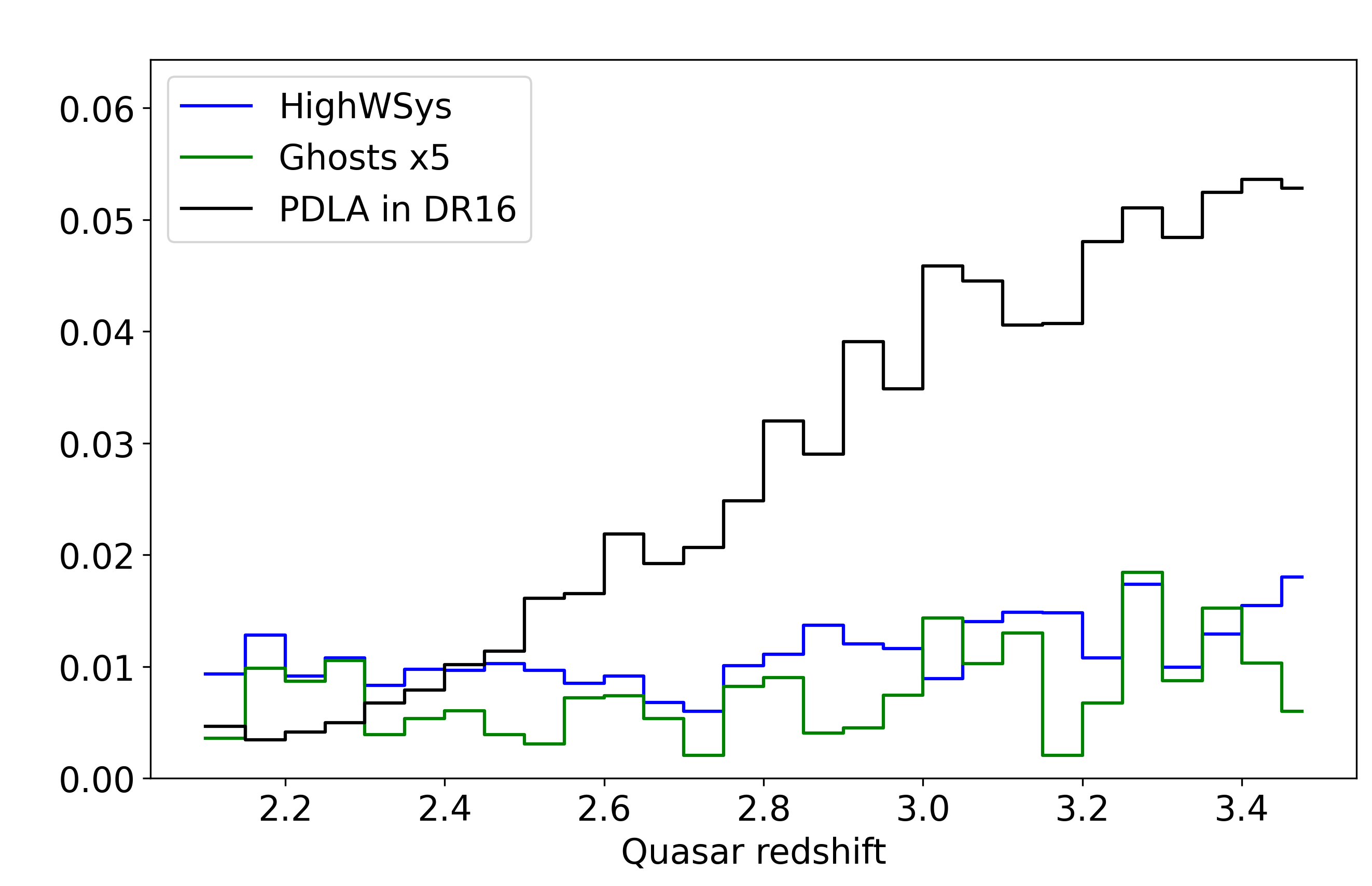}
      \caption{Ratio of the number of PDLAs in \cite{chabanier2022} (black),
      HighWSys (blue) and Ghost+Ghost? (green) to the total number of 
      DR16 quasars with the same selection criteria. The ratio is multiplied
      by a factor of five for Ghost DLAs.}
         \label{Distributionz}
   \end{figure}

In Fig.~\ref{Distributionz} we plot the distribution of the ratio of HighWSys (blue), Ghost (including probable Ghost) (green) and PDLAs in DR16 (black) from \cite{chabanier2022} 
to the number of quasars in DR16 (with the same selection criteria) 
versus redshift. The ratio is multiplied by a factor of five for Ghost DLAs.
It is striking to note that although the number of PDLAs increases 
with redshift, the number of systems detected in this work does not. 
The ratio of our systems to the total number of quasars is remarkably constant.
The surprising low number of PDLAs detected at $z<2.4$ may be due to the 
fact that it is more difficult to detect them at the blue end of the spectrum
whereas we do not have this difficulty for metal line systems. 
This could be a consequence of the presence of much more gas with low metallicity
at high redshift but more probably a consequence of a mixture of all this.
Note also that this may 
explain in part the large number of our systems that are not in the PDLA sample.

\section{Median spectra}
In this section, we derive a median spectrum for several of the system classes defined above, allowing us to distinguish them on a physical basis. To do so, we stack the spectra of the corresponding quasars after shifting them to the rest frame of the absorbers.

\begin{figure*}[htbp]
\centering
\includegraphics[width=0.8\textwidth,height=0.5\textheight
]{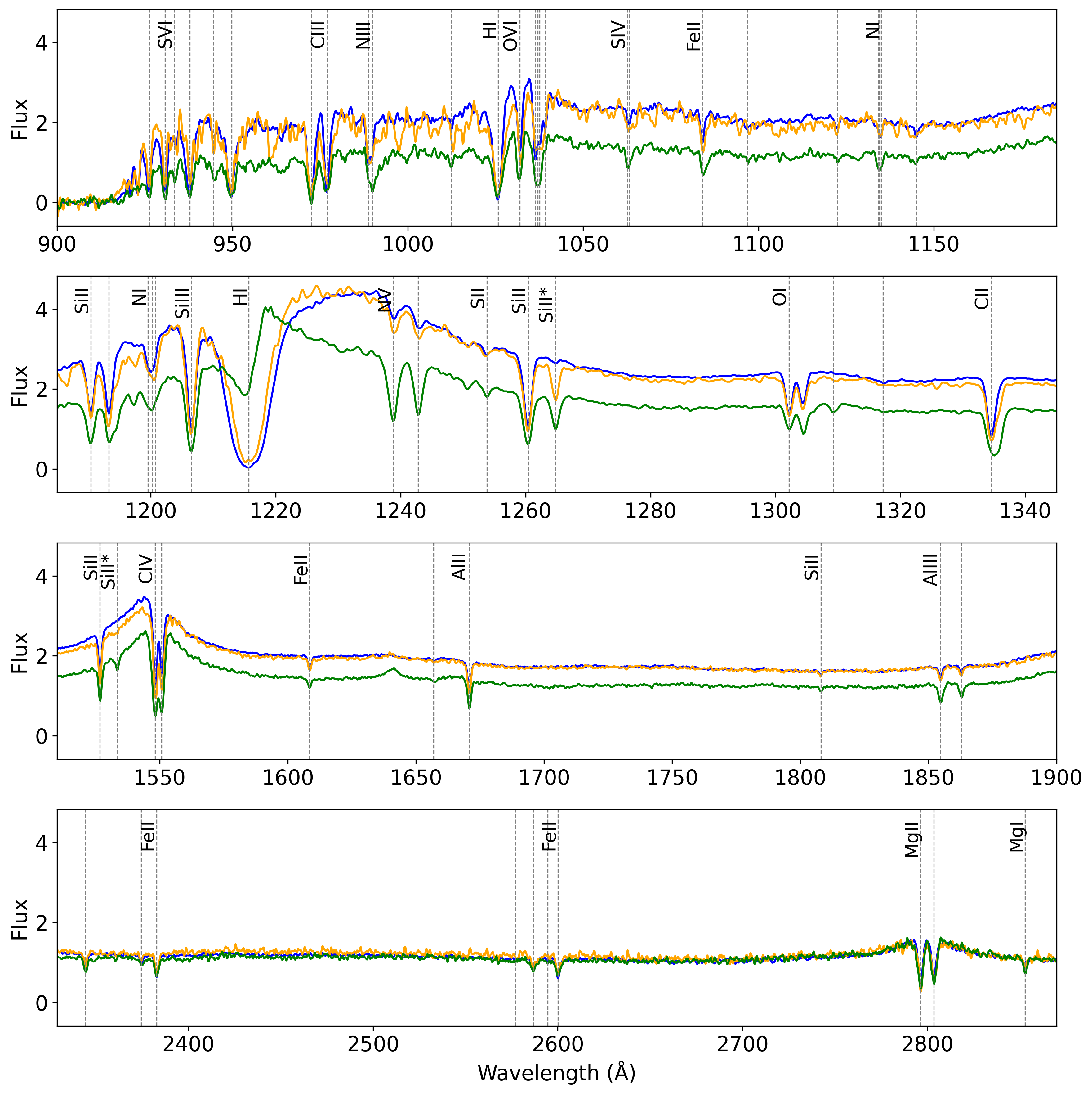}
\caption{Median spectra of the Ghost+Ghost? (green), HighWSys (blue) and 
HighWSys with Si~{\sc ii}$^*$ (orange)}
\label{fig:DLAAjoutsGhostSpectre}
\end{figure*}

For the Ghost and Ghost? systems, we obtained the median spectra separately and found them to be very similar, except that the H {\sc i} absorption is slightly more pronounced in the Ghost? class. We therefore combined all Ghost and Ghost? systems into a single sample.
In Fig.~\ref{fig:DLAAjoutsGhostSpectre}, we show the median spectrum of the combined Ghost+Ghost? sample in green together with that of the HighWSys in our sample in blue. Both spectra are normalised within a 100~\AA~ window around 2900~\AA.

\begin{figure}[h]
   \centering
   \includegraphics[width=0.8\linewidth, keepaspectratio]{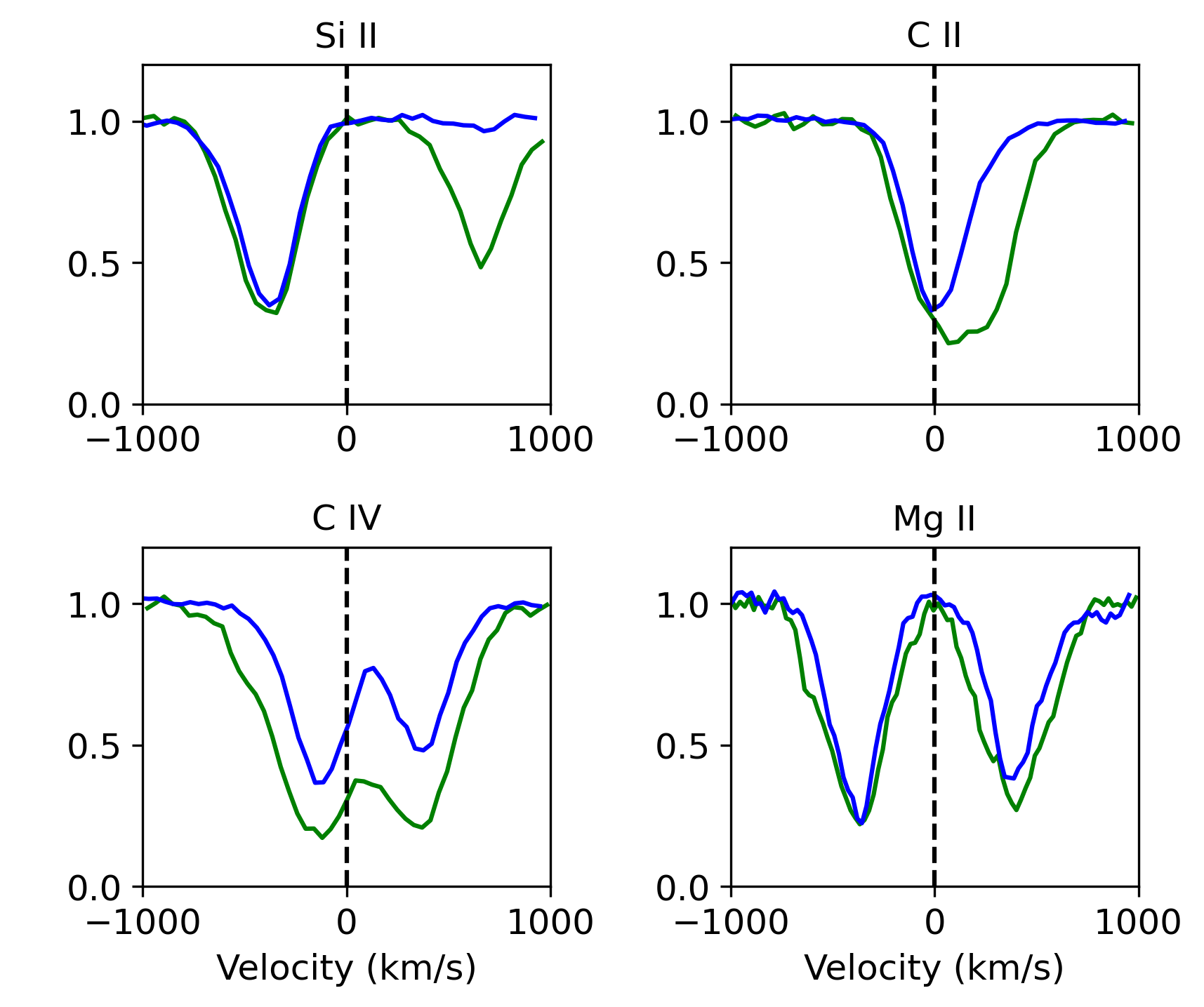}
      \caption{Portions of the normalised median spectra of HighWSys (blue) and Ghost DLAs (green) around a few transitions. Upper panels: Si{\sc ii}$\lambda$1260 and Si{\sc ii}$^*$$\lambda$1264 on the left, C{\sc ii}$\lambda$1334 and C{\sc ii}$\lambda$1335 on the right. Lower panels: C{\sc iv} doublet on the left and Mg{\sc ii} doublet on the right.}
         \label{Vitspecies}
   \end{figure}

It is apparent that the Ghost spectrum is more attenuated, suggesting a larger dust content in Ghost DLAs. In addition, the absorption lines arising from excited levels of Si {\sc ii} 
and C {\sc ii} are significantly stronger in Ghost DLAs. Although most absorption lines exhibit similar strengths, the Mg {\sc ii} and C {\sc iv} profiles are noticeably broader in Ghost DLAs 
(see Fig.~\ref{Vitspecies}). Moreover, the absorptions in the Lyman series are strong clearly indicating that Ghost DLAs are bona fide DLA systems.

During the visual inspection, we flagged all systems in which absorption from excited levels of Si~{\sc ii} is detected. In Fig.~\ref{fig:DLAAjoutsGhostSpectre}, we overplot the median spectrum of these systems on top of the HighWSys spectrum. The strengths of the Si~{\sc ii}$^*$ and N~{\sc v} absorptions
suggest a sequence among the three spectra, indicating exposure to a more intense UV radiation field.

\begin{figure*}[htbp]
\centering
\includegraphics[width=0.8\textwidth,height=0.4\textheight
]{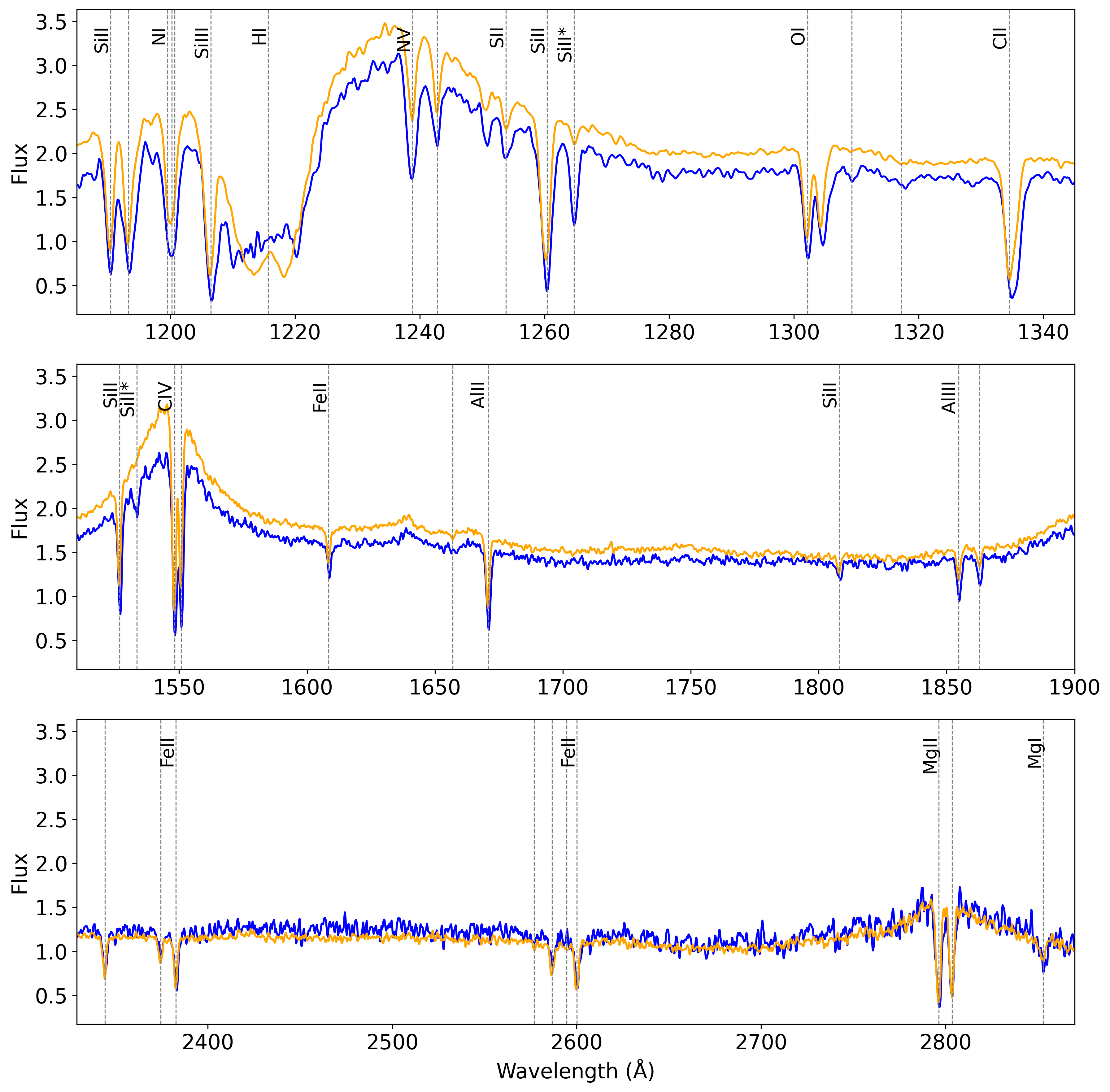}
\caption{Median spectra for the DLA-Cor with (blue) and without (orange) Si~{\sc ii}$^*$.}
\label{fig:DLACorSpectre}
\end{figure*}

In Fig.~\ref{fig:DLACorSpectre}, we show the median spectra of the DLA-Cor systems with (blue) and without (orange) detectable Si~{\sc ii}$^*$. A narrow emission line is visible in the DLA trough for the systems without (or with only weak) Si~{\sc ii}$^*$ absorption, whereas the systems with strong Si~{\sc ii}$^*$ 
absorption display a trough which is not going to the zero level but has
no clear narrow emission line. The DLA-Cor* spectrum is also slightly
more attenuated.
This strongly suggests that the systems with detectable 
Si~{\sc ii}$^*$, DLA-Cor*, represent a transition between DLA-Cor and Ghost DLAs.  The absence of a narrow emission line is likely a consequence of the larger kinematic spread of the DLA-Cor* systems around the QSO redshift, which washes out the emission line.

The number of systems flagged for the presence of Si~{\sc ii}$^*$ 
absorption are 54 HighWSys (out of 1688), 101 LowWSys (out of 1554),
17 VeryLowWSys (out of 119) and 48 DLA-Cor (out of 227). 
Therefore, the peculiar systems (either not covering the whole
emission region, DLA-Cor+DLA-Cor?+Ghost+Ghost? or being associated with excited species or both)
represent 583 out of 3774 objects (HighWSys+LowWSys+VeryLowWSys in addition
to the particular objects) thus 15\%. 

\begin{figure*}[htbp]
\centering
\includegraphics[width=0.8\textwidth,height=0.4\textheight
]{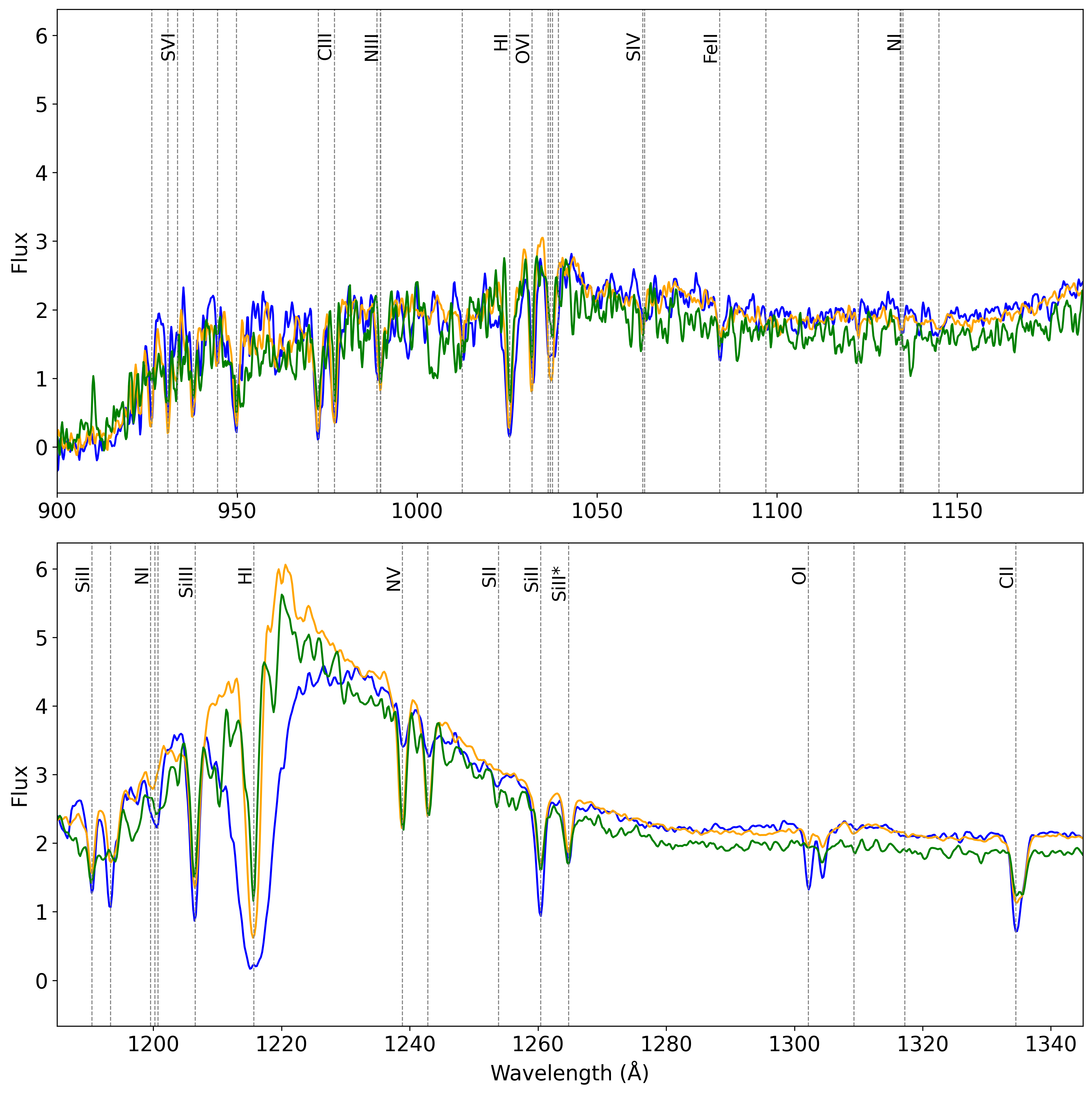}
\caption{Median spectra for the HighWSys (blue), LowWSys (orange) 
and VeryLowWSys (green) with Si~{\sc ii}$^*$ absorption detected.}
\label{fig:AllAjoutsSpectre}
\end{figure*}

In Fig.\ref{fig:AllAjoutsSpectre} we plot part of the spectrum of 
these systems (HighWSys, LowWSys and VeryLowWSys) with Si~{\sc ii}$^*$ detected.
It can be seen that (i) the Si~{\sc ii}$^*$ absorptions are
very similar although the H~{\sc i} absorptions are very 
different; (ii) the C~{\sc ii}$^*$/C~{\sc ii} ratio but also the N~{\sc v} absorptions are stronger at lower $N$(H~{\sc i}). 
We note that there is a small signature of dust attenuation as for Ghostly DLAs 
(see Fig.~\ref{fig:DLAAjoutsGhostSpectre}).
It seems that the sub-DLAs* are closer to the central AGN
and that the reduced H~{\sc i} column density is
a consequence of ionisation.

\begin{figure*}[htbp]
\centering
\includegraphics[width=0.8\textwidth,height=0.4\textheight
]{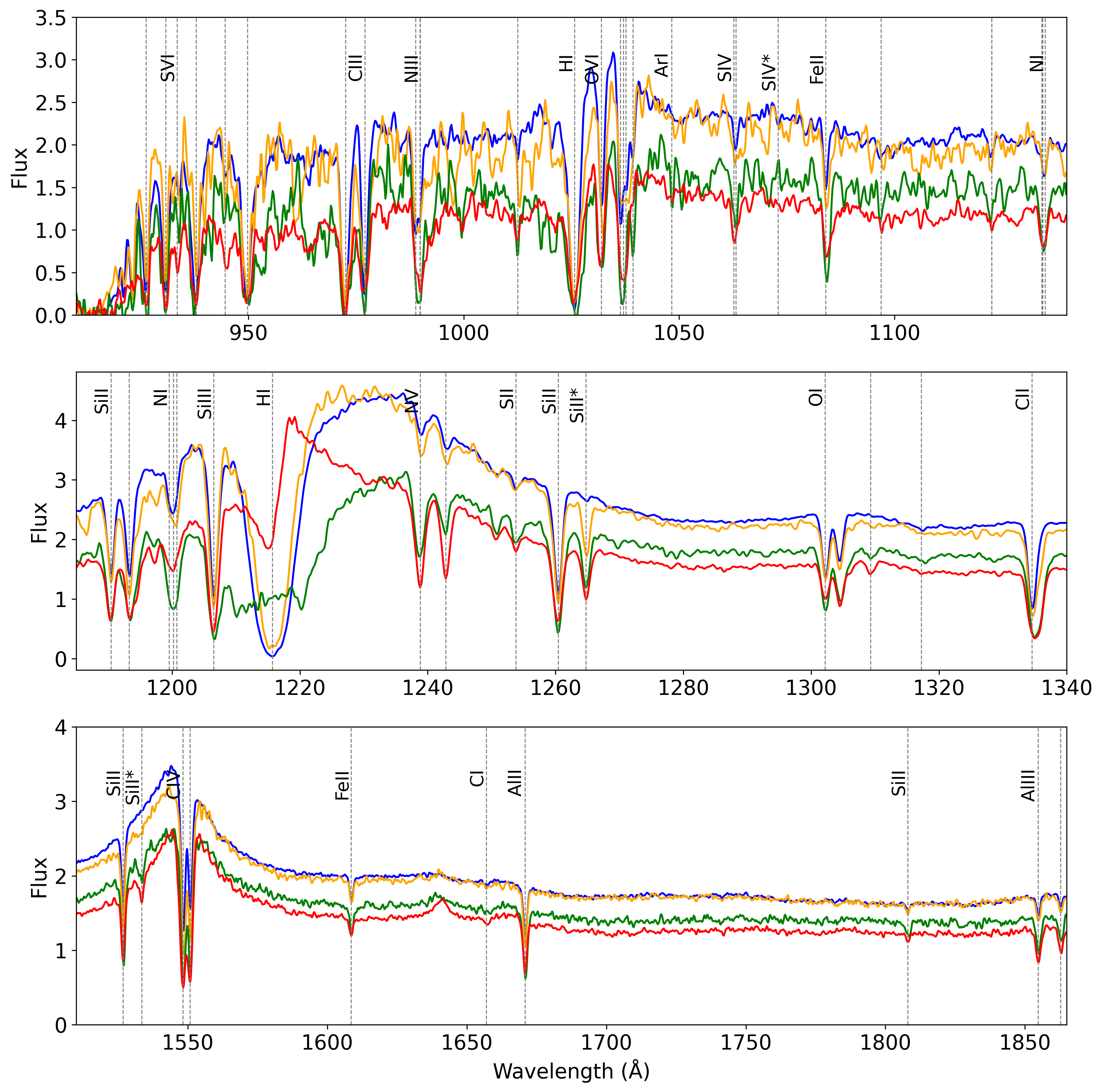}
\caption{Median spectra for the HighWSys (blue), HighWSys with Si~{\sc ii}$^*$ (orange), 
DLA-Cor with Si~{\sc ii}$^*$ (green) and Ghostly DLAs (red).}
\label{fig:sequence}
\end{figure*}

Finally we compare in Fig.\ref{fig:sequence} the spectra of 
HighWSys (blue), HighWSys with Si~{\sc ii}$^*$ (orange), 
DLA-Cor with Si~{\sc ii}$^*$ (green) and Ghostly DLAs (red)
It is apparent that there is a sequence  HighWSys/HighWSys*/DLA-Cor*/Ghost
with higher excitation and ionisation from DLAs to Ghosts
revealed by stronger Si~{\sc ii}$^*$ absorption and 
C~{\sc ii}$^*$/C~{\sc ii} ratio plus stronger C~{\sc iv}, N~{\sc v} and O~{\sc vi} absorptions \citep[see also][]{Balashev2023}.

\section{Velocity distributions}
\begin{figure}[h!]
    \centering
        \includegraphics[width=0.9\columnwidth]{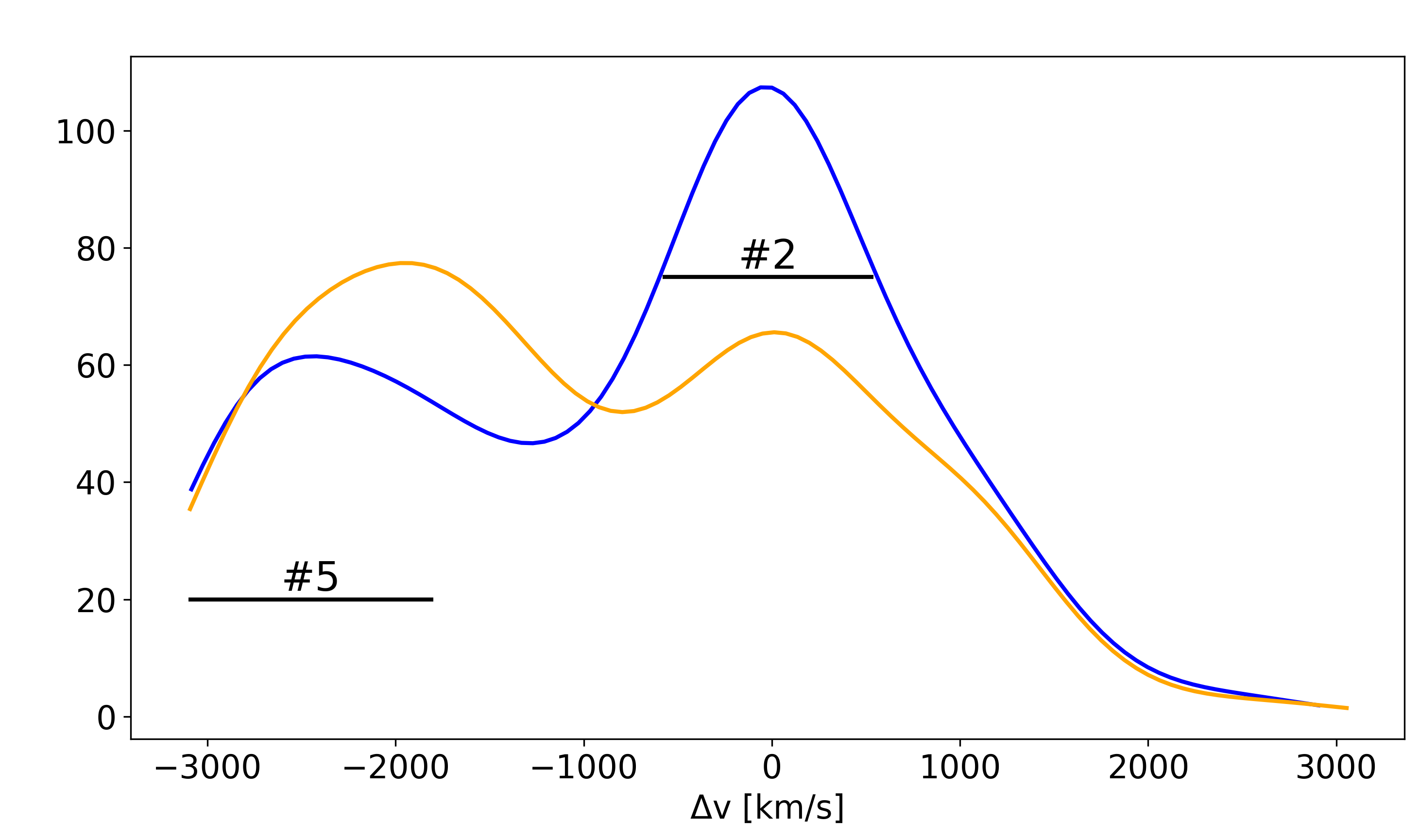} 
        \includegraphics[width=0.9\columnwidth]{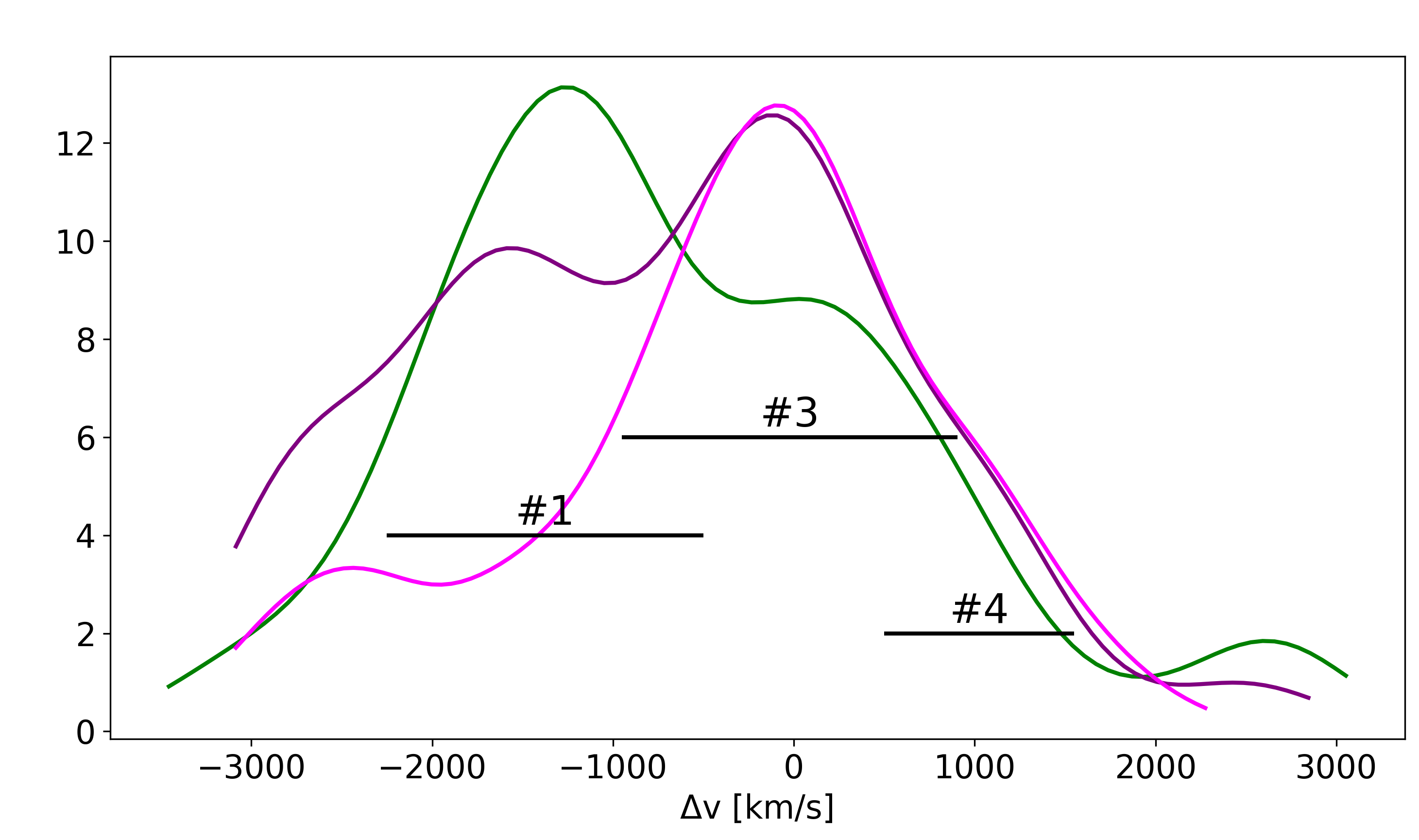}
 
    \caption{Distributions of the velocity relative to the quasar for, in the upper panel, HighWSys (blue) and LowWSys (orange), and in the bottom panel, Ghostly DLAs (green), all systems with detected Si~{\sc ii}$^*$ absorption (purple) and DLA-Cor without any Si~{\sc ii}$^*$ absorption (magenta). The regions \#1 to \#5 are discussed in Section~7.}
    \label{fig:velocities}
\end{figure}

We study here the kinematics of different classes of absorbers to determine whether this parameter allows us to distinguish between them. In Fig.~\ref{fig:velocities}, we show two histograms smoothed using Kernel Density Estimation, representing the distributions of velocities relative to the quasar emission redshift for the different classes.

In the top panel, the blue and orange curves correspond to HighWSys and LowWSys, respectively, with about 1500 objects in each sample. In both cases, two distinct regions are visible.
The peak centred near zero velocity corresponds to systems located close to the quasar redshift, likely associated with the host galaxy. The width of this peak, approximately 1400~\kms\ full width at half maximum (FWHM), is larger than the uncertainty in quasar redshift, which we estimate to be between 700 and 1000~\kms\ \citep{Paris2018}. This indicates that the neutral gas associated with the galaxy is highly turbulent.
There is a minimum in the distributions around $-1200$~\kms, beyond which the distribution rises again. This likely corresponds to a deficit of systems caused by the clearing of neutral gas around the galaxy by the strong UV flux. At velocities smaller than $-2000$~\kms, we observe the onset of the intervening DLA population.

In the bottom panel, we plot the same distributions for Ghostly DLAs (green curve), all systems where Si~{\sc ii}$^*$ absorption is detected (purple curve), and DLA-Cor systems with no Si~{\sc ii}$^*$ absorption (magenta curve), with approximately 200 objects in each sample. These systems correspond to peculiar absorbers identified above, likely located in the central part of the host galaxy, where some clouds either do not fully cover the background emission source or where silicon is excited by the physical conditions characteristic of these regions.
Two distinct regimes are apparent. A peak around the quasar redshift, similar to that seen in the top panel, is present. However, in this case, the peak is broader, with a full width at half maximum of about 2200~\kms. This is the only peak in the distribution of DLA-Cor systems without excited Si~{\sc ii} absorption. In contrast, the other two distributions show a second peak around $-$1000 to $-$2000~\kms. Most Ghost DLAs have such velocities, coinciding with the lack of simple DLAs
at these velocities as seen in the
top panel. This strongly
suggests that these Ghostly DLAs are part of an AGN-driven outflow.
We interpret these observations in Section~7.

\section{Quasar properties}
\citet{Wu2022} produced a catalogue of continuum and emission-line properties for broad-line quasars in the SDSS DR16 catalogue \citep{Lyke2020}. Here, we explore the characteristics of the quasars with ProxSys systems using their results.

\begin{figure}[h]
   \centering
\includegraphics[width=0.9\columnwidth]{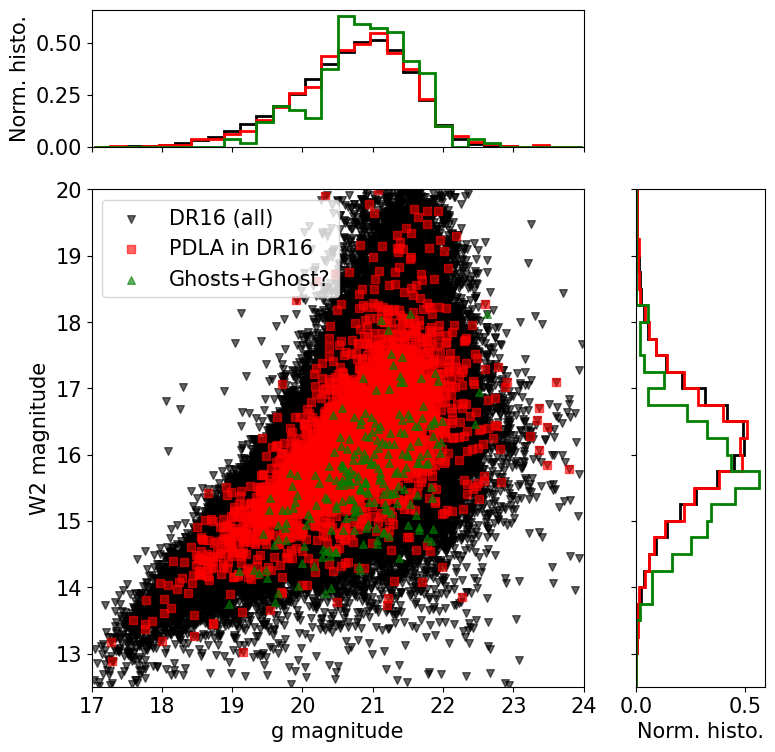}
      \caption{g magnitude versus W2 magnitude for all DR16 quasars
      (black), quasars with PDLA from the sample of \citet{chabanier2022} (red) and quasars with Ghost+Ghost? of this work (green). The histograms of g and W2 are represented on the right and top axis.}
         \label{gvsW2}
   \end{figure}

Fig.~\ref{gvsW2} shows the g-band versus W2-band magnitudes for quasars in DR16 (black), quasars with PDLAs from the sample of \citet{chabanier2022} (red), and quasars with Ghost or Ghost? systems from this work (green). 
It is apparent that quasars with Ghost+Ghost? systems are brighter in W2 while exhibiting similar g-band magnitudes. The median values of g and W2 magnitudes for the three samples are 20.766, 20.944, 20.914 and 16.167, 16.228, 15.613, respectively.

\begin{figure}[h]
   \centering
\includegraphics[width=0.9\columnwidth]{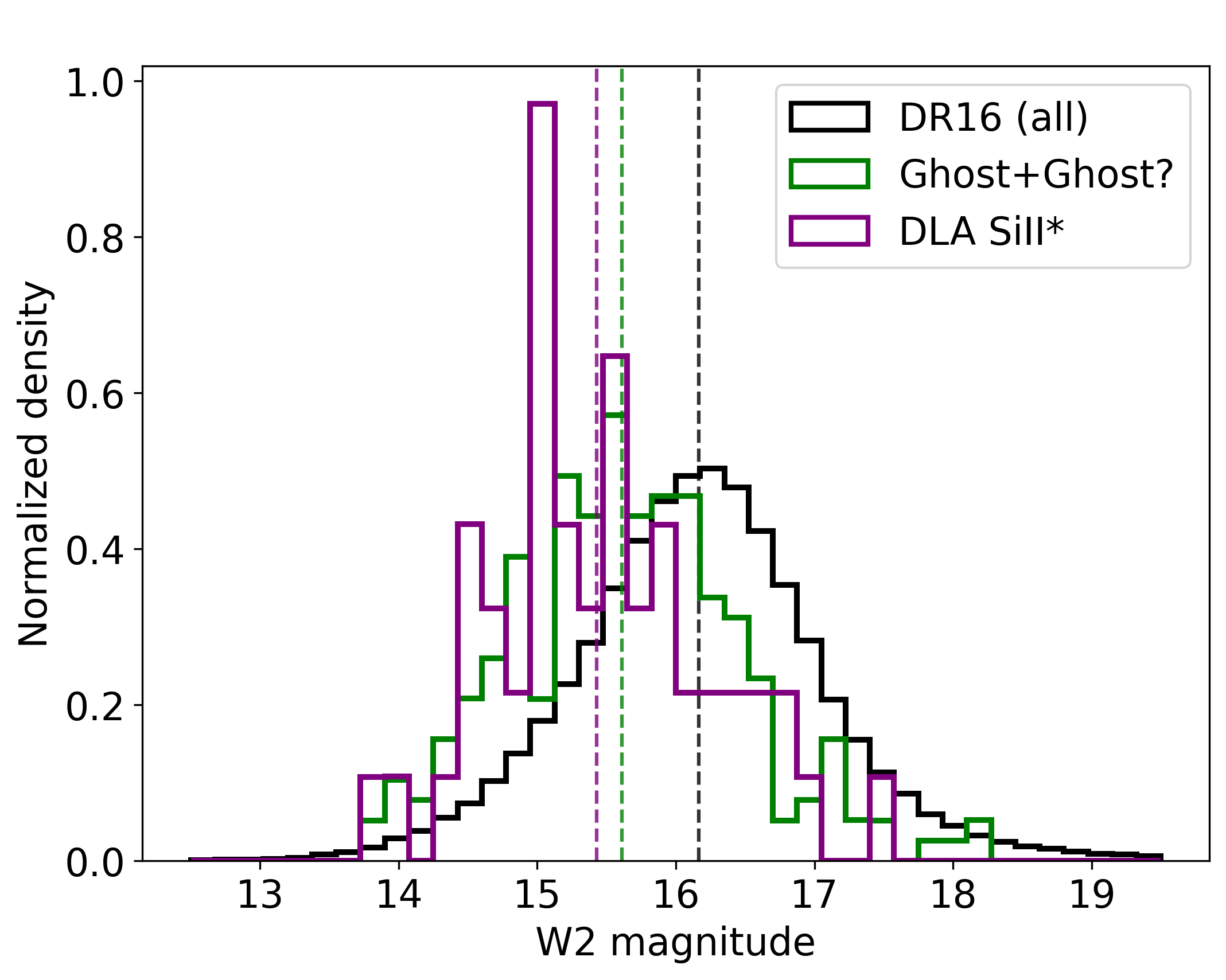}
      \caption{Normalised histogram of W2 magnitude in the samples of DR16 quasars (black), quasars with Ghost+Ghost? systems (green) and quasars with DLAs showing Si~{\sc ii}$^*$ absorption (purple).}
         \label{HistoW2}
   \end{figure}

Interestingly, quasars in which a DLA with Si~{\sc ii}$^*$ is detected are also brighter in W2, as shown in Fig.~\ref{HistoW2}. The median W2 magnitude for this sample is 15.431.

\begin{figure}[h]
   \centering
 
\includegraphics[width=0.9\columnwidth]{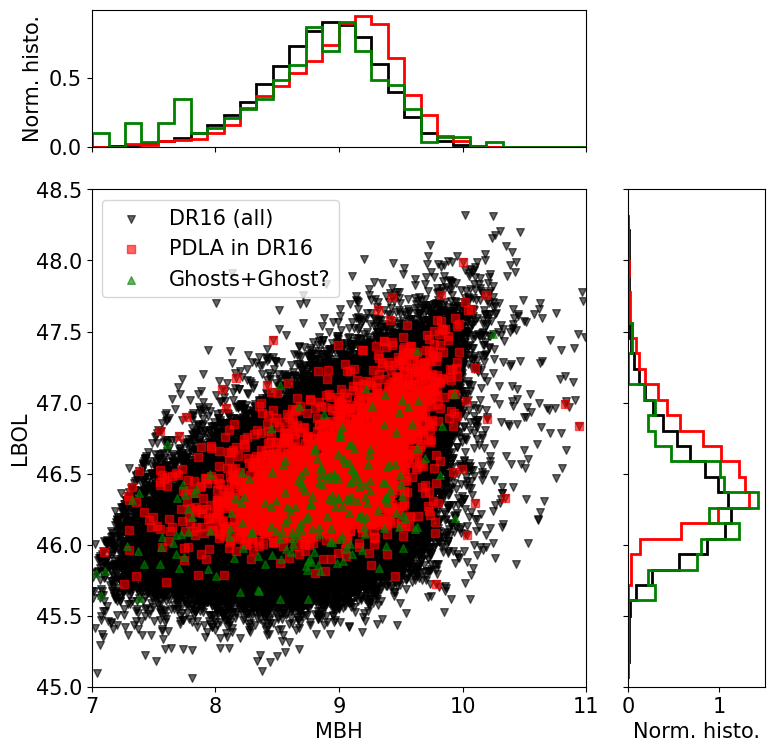}
      \caption{Bolometric luminosity versus black-hole mass for all DR16 quasars (black), quasars with PDLA from the sample of \citet{chabanier2022} (red) and quasars with Ghost+Ghost? of this work (green). The normalised histograms of LBOL and MBH are represented on the right and top axis.}
         \label{LBOLvsMBH}
   \end{figure}

Fig.~\ref{LBOLvsMBH} shows the bolometric luminosity versus black-hole mass for quasars in DR16 (black), quasars with PDLAs from the sample of \citet{chabanier2022} (red), and quasars with Ghost+Ghost? systems from this work (green). The median values of $M_{\rm BH}$ and $L_{\rm bol}$ for the three samples are 8.89, 9.05, 8.85 and 46.32, 46.50, 46.30, respectively.

PDLA quasars are slightly more luminous and host marginally more massive black holes. Interestingly, we consistently recover these differences if we replace the PDLA sample from DR16 with our HighWSys sample. This could reflect the greater prevalence of systems with large velocity offsets relative to the quasar (and thus likely associated with companion galaxies), as seen in Fig.~2, if brighter quasars tend to reside in richer environments \citep[see also][]{Ellison2010}.

\begin{figure}[htbp]
\centering
\includegraphics[width=\textwidth,height=0.2\textheight,keepaspectratio]{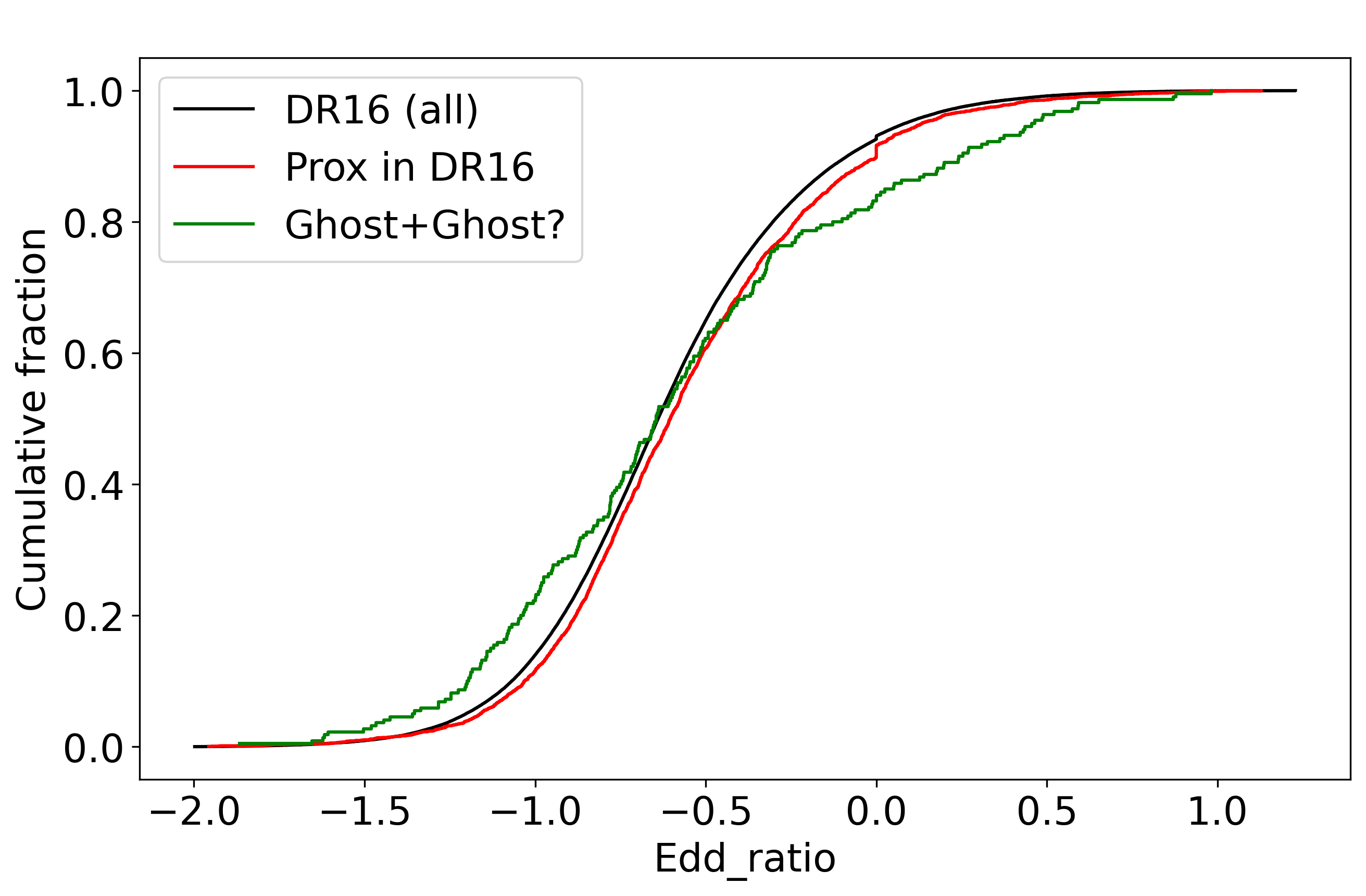}

\caption{Cumulative distribution of Eddington ratios for quasars in DR16 (black), DR16 PDLAs (red) and Ghost+Ghost? (green). 
}
\label{fig:Eddcum}
\end{figure}

Although direct evidence of feedback is difficult to obtain, it has long been suggested that feedback is somehow related to the Eddington ratio \citep[e.g.,][]{fabian2009}. Fig.~\ref{fig:Eddcum} shows the cumulative distribution of the Eddington ratio for the same three samples.  
The KS test yields a p-value of 0.018 and 0.0019 between 
the Ghost distribution and the DR16 quasars and the DR16 PDLAs distributions, respectively, indicating that the distributions are significantly different at the 5\% level.

The difference between DR16 quasars and quasars with PDLAs may be explained by the same effect discussed above. More surprisingly, however, is the difference between DR16 quasars and quasars with Ghost+Ghost? systems: there is an excess of the latter at both low and high Eddington ratios as ascertained by an Anderson-Darling test with p-value smaller than 10$^{-5}$. This will be discussed further in Section~\ref{sec:discussion}.

\begin{figure}[htbp]
\centering
\includegraphics[width=\textwidth,height=0.21\textheight,keepaspectratio]{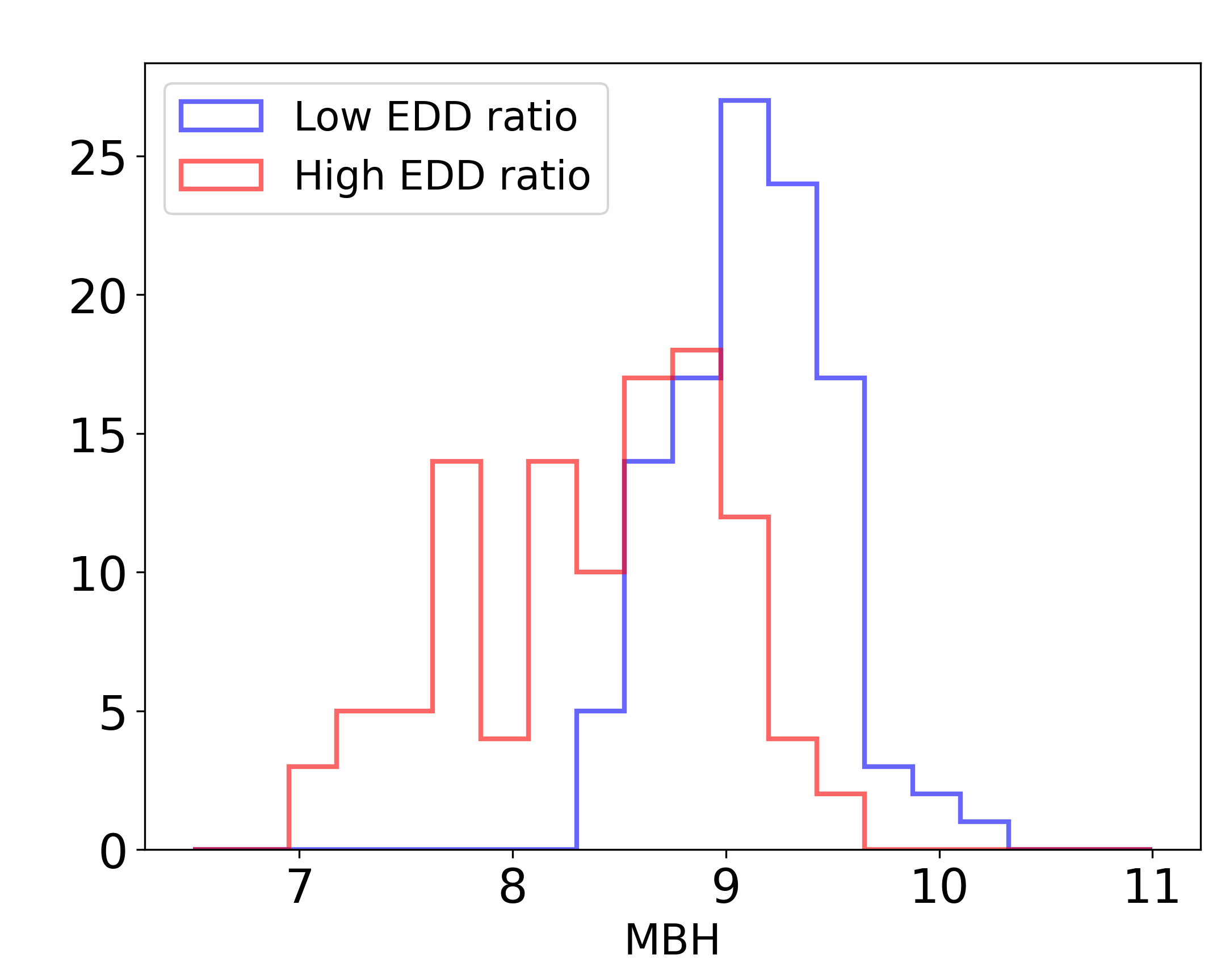}
\vspace{0.1cm}
\includegraphics[width=\textwidth,height=0.21\textheight,keepaspectratio]{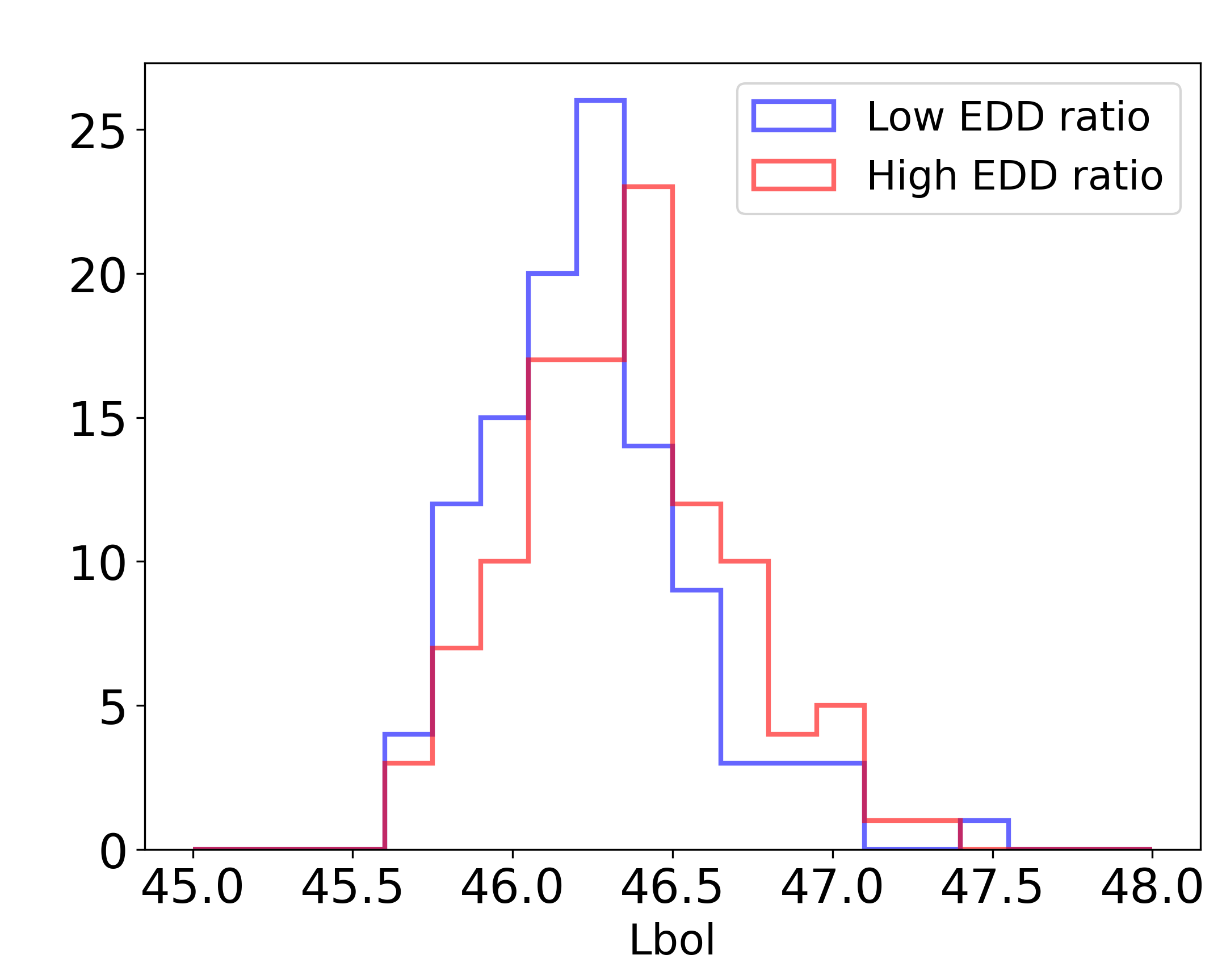}
\vspace{0.1cm}
\includegraphics[width=\textwidth,height=0.21\textheight,keepaspectratio]{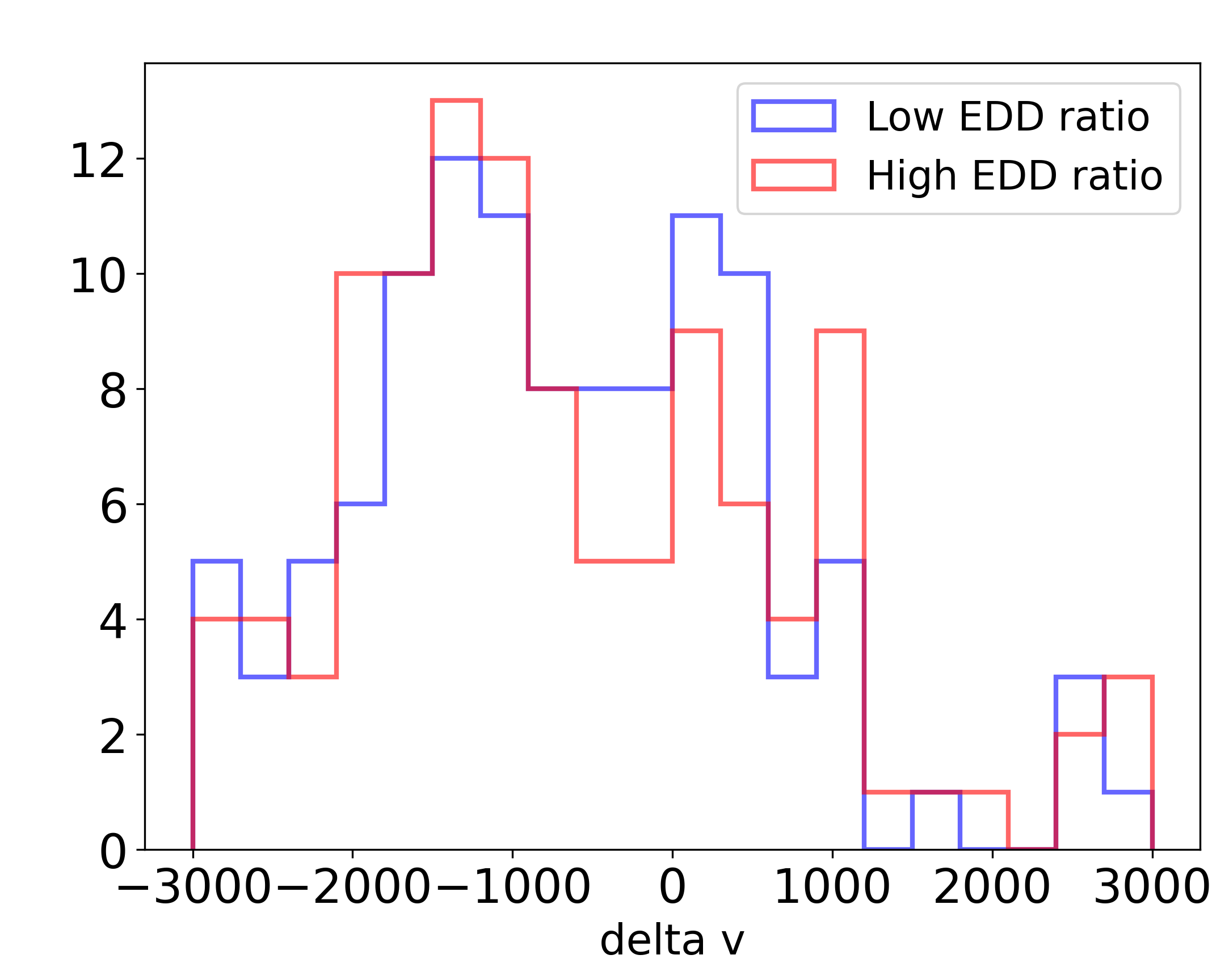}
\vspace{0.1cm}
\caption{Top panel: Black-hole mass distributions of quasars with Ghost+Ghost? systems, split into low (red) and high (blue) Eddington-ratio subsamples.
Middle panel: Same for the bolometric luminosity. Bottom panel: Same for the velocity of the systems relative to the quasar.
}
\label{fig:lowhigheddratioghost}
\end{figure}

To illustrate this within the Ghost+Ghost? sample, Fig.~\ref{fig:lowhigheddratioghost} shows the distributions of black-hole mass (top panel), bolometric luminosity (middle panel), and velocity relative to the quasar (bottom panel) for the two sub-samples with low and high Eddington ratios. Each sub-sample contains 92 systems. Clearly, differences in the Eddington ratio are dominated by black-hole mass, with no significant difference in the velocities between the two sub-samples. Note that the Eddington ratio describes the state of the quasar at the time of observation, whereas the velocity of the cloud reflects processes occurring on a longer timescale. 

\begin{figure}[h]
   \centering
 
\includegraphics[width=0.9\columnwidth]{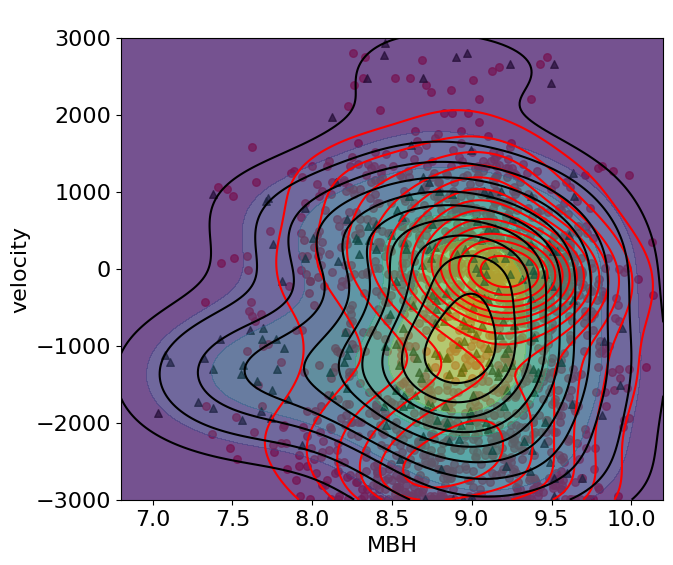}
      \caption{Velocity of systems versus black-hole mass. Black triangles and contours represent Ghost+Ghost? systems, while red points and contours correspond to HighWSys in our sample }
         \label{fig:velocityvsMBH}
   \end{figure}

We explored the possible correlation between velocity and the characteristics of the quasars. No clear trend was observed, except for a slight dependence on black-hole mass, as illustrated in Fig.~\ref{fig:velocityvsMBH}. Black triangles and contours represent Ghost+Ghost? systems, while red points and contours correspond to HighWSys in our sample. There appears to be a weak tendency for smaller black-hole masses to be associated with smaller velocities in the Ghost sample. 
There is a statistically significant shift in the location of the KDE maximum ($\Delta {\rm MBH}$ =  $-0.28 \pm 0.07$
and $\Delta {\rm V}$ = $-989 \pm 286$).
Interestingly, Ghost+Ghost? systems are distributed across all black-hole masses, whereas DLAs occur less frequently at lower masses.

\section{Discussion}
\label{sec:discussion}
The distributions of the system velocities relative to the quasar redshift 
(Fig.\ref{fig:velocities}), show different components:
(i) an outflow component (component \#1) with velocities up to 
$\sim$$-$2200~\kms~ mostly
seen in Ghostly systems (small clouds with large density, see below);
(ii) two components centred on the quasar redshift: 
one with smaller 
dispersion (component \#2 seen for HighWSys and LowWSys) originating 
in gas in the host galaxy, and 
one with larger dispersion (Ghostly DLAs, systems with Si~{\sc ii}$^{*}$ and
DLA-Cor, component \#3) a component associated with the host galaxy but with a 
disturbed kinematics which might correspond to the final fate of the outflow component ;
(iii) one component (\#4) with positive velocity up to 1800~\kms~ which is
associated with gas bouncing back to the centre of the quasar and 
(iv) a component with large negative velocities (\#5) corresponding to 
systems associated with companions of the quasar host galaxy.
In the following we will discuss kinematic components \#1 to \#4.
Component \#5 corresponds to the standard interpretation of
PDLAs and is clearly identified. 

\par\noindent
\subsection{Component \#1 and Ghostly systems}
A large fraction of the Ghostly systems are apparently part of 
an outflow with
typical velocities in the range $-$1000 to $-$2000~\kms~ and one can
conjecture that the gas in these systems originates from 
above the disc of the host galaxy, may be the dusty torus, and is driven by radiation or central disc wind. 

A detailed study of the physical characteristics of these systems from SDSS data is difficult, if not impossible, as column densities cannot be robustly derived from the absorption features due to the low spectral resolution of the spectra
\citep[e.g.][]{Prochaska2006b}. The derivation of column densities from such data is a common practice that we will avoid here. We will devote future work to a more 
detailed characterisation of their properties.
We can however discuss some of these characteristics useful here.

The presence of Si~{\sc ii}$^*$ can be interpreted by collisional excitation.
Other processes, direct pumping by IR radiation or indirect pumping via UV excitation
like in the ISM of GRB host galaxies \citep{Vreeswijk2007} are unlikely, in particular
the absence of absorption from the excited level 
Fe~ {\sc ii}$^6$D$_{7/2}$ at $\lambda$2396 favours the collisional excitation origin \citep{Prochaska2006}.
From the spectra in Fig.\ref{fig:AllAjoutsSpectre} we can say 
that the 
Si~{\sc ii}$^*$/Si~{\sc ii} column density ratio is 
in the range 0.5, if the lines are optically thin, to 1, if they are
optically thick.  
From \citet{Silva2002}, given that collisions with electrons are the 
most effective,
we can derive (see their Fig.~8) that the electronic density is in the
range 200-1000~cm$^{-3}$ \citep[see also Fig.~6 of][]{Srianand2000}. 
Same range can be derived from the fact that the 
C~{\sc ii}$^*$$\lambda$1335 is stronger than the 
C~{\sc ii}$\lambda$1334 line (see Fig.~\ref{fig:sequence}).

We can estimate a lower limit on the distance of the Ghost gas 
from the central engine using the presence of Mg~{\sc i} absorption 
and following \citet{Prochaska2006}. Considering the gas in 
ionisation equilibrium, a Mg~{\sc ii} recombination coefficient of 
$\alpha$~=~10$^{-12}$~cm$^3$/s, a Mg~{\sc i} ionisation 
cross-section of $\sigma$~=~2$\times$10$^{-18}$~cm$^2$, for a ratio
Mg~{\sc i}/Mg~{\sc ii}~=~0.01, an electronic density of $n_{\rm e}$~=~500~cm$^{-3}$
and a flux at 7.6~eV of $F$~=~10$^{-17}$~erg/s/cm$^2$/\AA~ 
we can estimate that the distance to the quasar is larger than 1~kpc.
This is large and means probably that the structure of the cloud
is more complex. High spectral resolution observation are required
to have a better understanding of the physical properties of these
clouds.
Another but still very uncertain lower limit on the distance can be obtained by equating the heating 
rate to the cooling rate in the cloud using the constraint on the  
C~{\sc ii}$^*$/C~{\sc ii} ratio following \citet{wolfe2003}. Using this constraint, 
\citet{Fathivavsari2017} found a lower limit of 39~pc. 
The clouds are not located in the close vicinity of the 
central engine but are still in its proximity. 

If the H~{\sc i} column density is of order 10$^{21}$ cm$^{-2}$~ 
(Fathivavsari et al. 2017), then a characteristic size of the Ghostly clouds
would be smaller than 1~pc, explaining the partial coverage of the BLR.
Note that partial coverage does not prevent the gas from being far from the centre. 
Indeed partial coverage is observed frequently for associated systems and
BALs \citep[][]{Crenshaw2003} and corresponding distances of the
gas to the centre are sometimes claimed larger than 1~kpc 
\citep[][]{Borguet2013,Arav2018}.

Therefore, Component \#1 is most likely associated with an 
outflowing gas located at distances of the order of tens to hundreds 
of parsecs from the central engine. This interpretation is supported by its kinematics, the large Si~{\sc ii}$^*$/Si~{\sc ii} and C~{\sc ii}$^*$/C~{\sc ii} ratios, and the presence of strong associated N~{\sc v} and O~{\sc vi} absorption, indicating a multiphase medium.

\subsection{Component \#2}
This component corresponds to the standard PDLAs. These are DLAs 
with velocities relative to the quasar within -1000 to 1000 \kms.
The properties of such systems have been studied by 
\citet{Prochaska2008} and \citet{Ellison2010}. 
Large distances, up to several tens of kpc, between the corresponding gas
and the central engine are found using 
the Wolfe et al. (2003) method  \citep[see also][]{Rix2007}.
However, recent studies of similar PDLAs bearing molecular hydrogen
place the systems much closer to the central engine.

\citet{Noterdaeme2019} searched for proximate H$_2$-bearing DLAs in SDSS DR14 and confirmed 81 systems with velocity offsets relative to the quasar smaller than 3000~km~s$^{-1}$. This represents an excess by a factor of 4–5 compared to the incidence of similar intervening systems. A significant, although generally small, residual flux is observed in the core of the DLA 
Lyman-$\alpha$ trough for about half of the systems. Absorption from excited atomic levels is detected in at least ten of them.
\citet{Noterdaeme2023} investigated the physical and kinematic properties of eleven of these systems and found that their velocity offsets relative to the quasar are within 500~km~s$^{-1}$. These absorbers therefore belong to component~\#2 or \#3. They concluded that the gas is located in the immediate vicinity of the central engine and may be associated with inflowing and/or outflowing material. A similar conclusion was reached by \citet{Han2026} for a PDLA at $z=2.40$ towards Q2310$-$3358, for which they derived a particle density of $n_{\rm H} \sim 1000$~cm$^{-3}$ similar to the density derived
for Ghostly systems.

\subsection{Component \#3}
The systems in this component has velocities in the range
$-$1000 to 1000~\kms~ but peaked at zero velocity. They are either 
Ghostly DLAs or systems with Si~{\sc ii}$^{*}$ or
DLA-Cor. Their kinematic distribution is broader than the 
one of the standard PDLAs (\#2). They show significant
residual in the core of the DLA trough and absorption from
excited atomic levels. This indicates that density is high
and their typical size is consistently small. They are associated
with stronger N~{\sc v} absorptions (see Fig.~\ref{fig:sequence}). It is tempting 
to see this gas as a consequence of the interaction between the
outflow and the ISM of the host-galaxy. 

\subsection{Component \#4} 
It can be seen in Fig.~\ref{fig:velocities} that the velocity distribution of Ghostly systems extends beyond $+1000~\kms$.
These systems having the 
same characteristics as the other Ghostly systems must be of high
density and located within a few hundreds pc from the centre.
They are however inflowing. There is no clear difference in the spectra
of these systems compared to the spectra of other Ghostly systems.
This component is seen also for the other types of systems.

\citet{Gaspari2017} developed the so-called chaotic cold accretion (CCA) model, in which gas inflow occurs as cold clouds condense out of the hot medium and migrate towards the centre through inelastic collisions. These cold clumps
share similar physical characteristics with those inferred for component~\#4.
The accretion energy is subsequently converted into ultra-fast outflows that entrain warm gas outward. In this framework, the velocities of the warm gas (typically associated with component~\#1) can reach up to $\sim$1000~\kms. These clouds are expected to lie at distances of a few hundred parsecs from the nucleus and to exhibit large internal velocity dispersions, as observed in Ghostly systems \citep{Fathivavsari2017}. A region characterised by highly turbulent kinematics may then develop at slightly larger distances, possibly corresponding to component~\#3.

\subsection{Geometry}
We can interpret the fact that the Ghostly systems are 
seen in quasars with brighter W2 magnitude as a geometrical effect
if these clouds are located close to the disc of the host galaxy.
Indeed, in that case the lines of sight intercepting these clouds have a 
preferential view of the inner side of the torus where the hot dust is located.

We have found 189 systems (so-called Ghost+Ghost?) where
the Lyman-$\alpha$ trough is not or barely seen indicating that the
corresponding clouds have transverse sizes smaller than the 
BLR. We found as well $\sim$400 systems showing a noticeable presence
of Si~{\sc ii}$^*$. In total this is about 15\%~ of the 4000 systems we
flagged in our study or 0.25\%~ of the 164,900 selected quasars in 
DR16. If the covering factor inside the regions where these clouds reside is of the order of 2\%~ \citep{Gaspari2018} and if we assume the clouds are 
located close to the galactic disc, the equatorial angle of the region 
is 8.5 degrees for Ghostly DLAs and 17 degrees for the systems with
Si~{\sc ii}$^*$ detected.

\subsection{The excess of Ghostly systems at low and high accretion rate}
We have seen on Fig.~\ref{fig:Eddcum} that the distribution
of the accretion rate of Ghostly systems show an excess
at low and high accretion rates compared to the population of quasars. 
We could easily imagine that a high accretion rate increases the 
ambient UV flux. To maintain a large column density of neutral hydrogen, the density of cold condensations would need to increase. In addition, the level 
of turbulence in the gas would also increase, as would the probability for 
a given line of sight to intersect such condensations.
For low Eddington ratio this seems at first glance more 
difficult to understand. However, the low Eddington ratio
is related to higher BH masses (see Fig.~\ref{fig:lowhigheddratioghost}). 
Higher BH masses correspond to 
deeper potential wells and therefore more turbulence in the CCA
scenario. The probability of a given line of sight intersecting 
a Ghostly system is 
therefore larger at both extremes.
 
We should note that \citet{vivek2025} show that the BAL fraction is increasing both at high and low Eddington ratios.

\section{Conclusion}
PDLAs have long been considered to be associated with gas in companion galaxies of the quasar host galaxy. The underlying idea is that gas in the host galaxy itself should be preferentially ionised by the central ionising source \citep{Prochaska2008, Ellison2010}. Indeed, \citet{Hennawi2007} showed that the probability of detecting an absorber with a high neutral hydrogen column density is much lower along the line of sight to the quasar than in the transverse direction \citep[see also][]{Guimaraes2007, Perrotta2016}. They conclude that this is a consequence of absorbers along the line of sight being photo-evaporated by direct exposure to the quasar’s ionising radiation. However, their Fig.~6 shows that a cloud can survive within 
$\sim$10~kpc of the quasar provided its density exceeds 
$\sim10^{3}$~cm$^{-3}$, which is consistent with the density we infer from the presence of Si~{\sc ii}$^*$ absorption in at least part of these systems.

More recently, the detection of H$_2$-bearing clouds at the quasar redshift
\citep{Noterdaeme2019}, as well as neutral gas clouds with sizes smaller than the broad-line region \citep{Finley2013, Fathivavsari2017}, has cast doubt on the conclusion that most PDLAs cannot be related to the quasar itself. Here, we show that such neutral gas not only exists but can be naturally interpreted as part of a cycling process within the framework of chaotic cold accretion \citep[CCA,][]{Gaspari2017}. The ubiquitous presence of strong N~{\sc v} and O~{\sc vi} absorption in all the spectra shown in 
\Cref{fig:AllAjoutsSpectre,fig:DLAAjoutsGhostSpectre},
together with the detection of Si~{\sc ii}$^*$ absorption, further supports this interpretation.

We do not expect the gas in Ghostly systems to be in equilibrium.
The large dispersion of the H~{\sc i} absorption lines
($b$~$>$~100~km~s$^{-1}$) suggests the gas be part of
a flow in which condensations can appear under the 
action of thermal instabilities \citep{Gaspari2017} or 
in the process of the disruption of a molecular cloud
by an outflow \citep{Lauzikas2024}. 
A constraint of about 10$^5$~yr on the typical active galactic nucleus (AGN) phase lifetime is inferred from the time lag between the moment when the AGN central engine switches on and becomes visible in X-rays, and the time required for the AGN to photo-ionise a large fraction of its host galaxy \citep{Schawinski2014}. 
This is of the same order of magnitude as the 
expected photo-evaporation time of a Ghostly system and 
in line with the CCA scenario.

In summary, the kinematical structures we have unveiled suggest a 
scenario in which AGN-driven outflows interact with a turbulent, multiphase ISM rather than a smooth medium. The hot, energy-driven wind percolates through low-density channels while simultaneously compressing denser clouds. Some of these cold, overdense clumps are accelerated and become part of the outflow, while others survive but are not sufficiently accelerated and eventually fall back towards the nucleus, forming a “fountain” flow. The outflow also injects turbulence, driving shocks and shear that can both destroy some clouds and generate new cold overdensities through radiative cooling. As a result, AGN feedback in such a medium is highly anisotropic and cyclical, with simultaneous gas inflow and outflow, rather than a simple, spherical clearing of the ISM.

To gain a deeper understanding of the origin of the gas we observe as Ghostly DLAs, high-resolution spectral data should be obtained to constrain its physical conditions, in particular its chemical composition and dust content but also internal kinematic structure, density and temperature.

 
 
\begin{acknowledgements}
This work was supported in part by the Agence Nationale de la Recherche (ANR, France) under contract ANR-22-CE31-0009.
We thank Camille No\^us (Laboratoire Cogitamus) for inappreciable and often unnoticed discussions, advice and support.
Funding for the Sloan Digital Sky Survey IV has been provided by the Alfred P. Sloan Foundation, the U.S. Department of Energy Office of Science, and the Participating Institutions. 

\end{acknowledgements}
\bibliographystyle{bibtex/aa}
\bibliography{bibtex/Patrick}

\end{document}